\documentclass[aps,prd,preprintnumbers,11pt,superscriptaddress,nofootinbib,showpacs]{revtex4-1}
\usepackage{amsmath}
\usepackage{bm}
\usepackage{times}
\usepackage{braket}
\usepackage{color,graphicx}

\definecolor{Red}{rgb}{1.,0.,0.}

\definecolor{nicered}{rgb}{0.7,0.1,0.1}
\definecolor{nicegreen}{rgb}{0.1,0.5,0.1}

\usepackage[pdftex]{hyperref}
\hypersetup{colorlinks,citecolor=nicegreen,linkcolor=nicered}

\begin{document}
\title{Heavy and light scalar leptoquarks in proton decay}

\author{Ilja Dor\v sner} \email[Electronic address:]{ilja.dorsner@ijs.si}
\affiliation{Department of Physics, University of Sarajevo, Zmaja od Bosne 33-35, 71000
  Sarajevo, Bosnia and Herzegovina}

\author{Svjetlana Fajfer} \email[Electronic
address:]{svjetlana.fajfer@ijs.si} 
\affiliation{Department of Physics,
  University of Ljubljana, Jadranska 19, 1000 Ljubljana, Slovenia}
\affiliation{J. Stefan Institute, Jamova 39, P. O. Box 3000, 1001
  Ljubljana, Slovenia}

\author{Nejc Ko\v snik} 
\email[Electronic address:]{nejc.kosnik@lal.in2p3.fr}
\affiliation{Laboratoire de l'Acc\'el\'erateur Lin\'eaire,
Centre d'Orsay, Universit\'e de Paris-Sud XI,
B.P. 34, B\^atiment 200,
91898 Orsay cedex, France}
\affiliation{J. Stefan Institute, Jamova 39, P. O. Box 3000, 1001 Ljubljana, Slovenia}

\begin{abstract}
  We list scalar leptoquarks that mediate proton decay via
  renormalizable couplings to the Standard Model fermions. We employ a
  general basis of baryon number violating operators to parameterize
  contributions of each leptoquark towards proton decay. This then
  sets the stage for investigation of bounds on the leptoquark
  couplings to fermions with respect to the most current Super
  Kamiokande results on proton stability. We quantify if, and when, it
  is necessary to have leptoquark masses close to a scale of grand
  unification in the realistic $SU(5)$ and flipped $SU(5)$
  frameworks. The most and the least conservative lower bounds on the leptoquark
  masses are then presented. We furthermore single out a leptoquark
  without phenomenologically dangerous tree-level exchanges that might
  explain discrepancy of the forward-backward asymmetries in $t \bar
  t$ production observed at Tevatron, if relatively light. The same
  state could also play significant role in explaining muon anomalous
  magnetic moment.  We identify contributions of this leptoquark to
  dimension-six operators, mediated through a box diagram, and
  tree-level dimension-nine operators, that would destabilize proton
  if sizable leptoquark and diquark couplings were to be
  simultaneously present.
\end{abstract}

\pacs{14.80.Sv,12.10.-g, 12.10.Kt}
\preprint{LAL 12-111}
\maketitle

\section{Proton decay leptoquarks}
There has been a plethora of low-energy experiments capable of
leptoquark discovery thus far. These have generated ever more
stringent constraints on available parameter space for their
existence. See, for
example,~\cite{Aad:2011uv,Chatrchyan:2011ar,Abazov:2011qj,Collaboration:2011qaa,Aad:2011ch,Aad:2012cy}
for some of the latest results. There also exists a large number of
phenomenological studies of leptoquark signatures prompted primarily
by various effects they could generate in flavor
physics~\cite{Leurer:1993em, Leurer:1993qx, Carpentier:2010ue,
  Saha:2010vw,Dighe:2010nj}. We are interested in a particular subset
of scalar leptoquark states that are associated with proton decay. It
is well-known that there exists only a small number of these states
that can simultaneously violate baryon ($B$) and lepton ($L$)
numbers~\cite{weinberg1,wilczek,weinberg2}. The number of scalar
leptoquarks that can mediate proton decay at the tree-level is even
smaller~\cite{Dorsner:2009cu}.  Our aim is to present a comprehensive
classification of leptoquarks and address a role these have in proton
decay processes.

Scalar leptoquarks that mediate proton decay certainly represent
qualitatively new physics. Although the relevant operators associated
with exchange of these states can be studied from an effective theory
point of view, we prefer to trace their origins to a particular
unification scenario in order to expose their dependence on underlying
couplings. In fact we will study these states in two different
unification frameworks that correspond to the
$SU(5)$~\cite{Georgi:1974sy} and the flipped
$SU(5)$~\cite{DeRujula:1980qc,Georgi:1980pw,Barr:1981qv}, i.e., $SU(5)
\times U(1)$, embeddings of the mater fields. These two scenarios are
general enough to cover other possible embedding schemes.

Let us start by spelling out qualitative differences between the
scalar and vector, i.e., gauge boson, leptoquarks that mediate proton
decay at the so-called dimension-six ($d=6$) level. (The latter have
been studied much more extensively in the literature. See, for
example,~\cite{weinberg1,wilczek,weinberg2,DeRujula:1980qc,FileviezPerez:2004hn,Dorsner:2004xx,Dorsner:2004jj,Dorsner:2004xa,Nath:2006ut}.)
Firstly, vector leptoquarks comprise twenty-four states whereas the
scalar ones comprise eighteen (fifteen) states in case neutrinos are
Dirac (Majorana) particles. Secondly, whereas $SU(5)$ contains only a
half of all vector leptoquarks, the other half being in flipped
$SU(5)$, one can already find all possible proton-decay mediating
scalar leptoquarks in either $SU(5)$ or flipped $SU(5)$
framework. Hence, the scalar sector, although smaller, can potentially
yield much richer structure with respect to the gauge one. Thirdly,
the uncertainty in predictions for partial nucleon decay rates due to
the gauge boson exchange resides entirely in a freedom to choose
particular unitary rotations that need to be in agreement with
observed mixing parameters in the fermionic sector as gauge bosons
couple to matter with the gauge coupling strength. Scalar fields, on
the other hand, couple to matter through Yukawa couplings. This brings
additional uncertainties to potential predictions for relevant decay
rates.

The leptoquark states that simultaneously violate $B$ and $L$ quantum
numbers tend to mediate proton decay at tree-level and are therefore
taken to be very massive. However, we have investigated an $SU(5)$
grand unified theory scenario~\cite{Perez:2007rm} which resulted in a
setup with a set of light leptoquarks. Namely, motivated by the need
to explain anomalous events in $ t \bar t$ production at
Tevatron~\cite{cdfttbar,Abazov:2011rq}, we have found that a light
color triplet weak singlet scalar could contribute to $t \bar t$
production and explain the observed increase of the forward-backward
asymmetry~\cite{Dorsner:2009mq}. We have accordingly demonstrated that
the unification of the fundamental interactions is possible if that
set of light scalars is a part of the $45$-dimensional
representation~\cite{Dorsner:2009mq}.

In flavor physics, due to recent accurate measurements at Tevatron and
LHCb, the presence of new physics (NP) in B systems seems rather
unlikely. The muon anomalous magnetic moment, on the other hand, still
leaves some room for NP contributions. The impact of potentially light
leptoquark scalars, including the light color triplet weak singlet
scalar, on the low energy and hadron collider phenomenology within
that context has been investigated in
Refs.~\cite{DelNobile:2009st,Vecchi:2011ab,Dorsner:2011ai,Dorsner:2010cu,Dorsner:2009mq}.

The color triplet weak singlet scalar state we have singled out does not
generate proton decay at the tree-level. However, one can still construct, as we show later, higher
order loop diagrams that yield effective $d=6$ and tree-level $d=9$
operators which can destabilize proton. The natural question then is
whether one can simultaneously address the $t \bar t$ asymmetry and
the muon anomalous magnetic moment by using the very same
leptoquark. We investigate this issue in detail in
Sec.~\ref{Higher_Order}.

This paper is organized as follows. In Sections~\ref{SU(5)} and
\ref{flippedSU(5)} we list all proton decay inducing leptoquarks in
$SU(5)$ and flipped $SU(5)$ unification frameworks and specify their
Yukawa couplings to the SM fermions. In Sec.~\ref{proton_decay} we
introduce the effective dimension-six operators for proton decay and
calculate associated effective coefficients for each leptoquark
state. Sec.~\ref{Discussion} is devoted to a study of conservative
lower bounds on the color triplet leptoquark mass within
phenomenologically realistic $SU(5)$ and flipped $SU(5)$ scenarios. In
Sec.~\ref{Higher_Order} we study leptoquarks that do not contribute to
proton decay operators of dimension-6 at tree-level. We conclude in
Sec.~\ref{Conclusions}.

\section{Leptoquarks in $SU(5)$}
\label{SU(5)}
The scalars that couple to matter at tree-level reside in the $5$-, $10$-, $15$-, $45$- and $50$-dimensional representations of $SU(5)$ because the SM
matter fields comprise $\bm{10}_i$ and $\overline{\bm{5}}_j$, where $i,j=1,2,3$ represent family indices. 
Namely, $\bm{10}_i=(\bm{1},\bm{1},1)_i \oplus(\overline{\bm{3}},\bm{1},-2/3)_i
\oplus(\bm{3},\bm{2},1/6)_i=(e^C_i,u^C_i,Q_i)$ and
$\overline{\bm{5}}_j=(\bm{1},\bm{2},-1/2)_j\oplus
(\overline{\bm{3}},\bm{1},1/3)_j=(L_j,d^C_j)$, where $Q_i=(u_i
\quad d_i)^T$ and $L_j=(\nu_j \quad e_j)^T$~\cite{Georgi:1974sy}. 
Possible contractions of the matter field representations hence read $\bm{10} \otimes \bm{10} =
\overline{\bm{5}} \oplus \overline{\bm{45}} \oplus \overline{\bm{50}}$, $\bm{10}
\otimes \overline{\bm{5}} = \bm{5}\oplus\bm{45}$ and $\overline{\bm{5}}
\otimes \overline{\bm{5}} = \overline{\bm{10}} \oplus \overline{\bm{15}}$. Theory also allows for addition of right-handed neutrinos that can be introduced, for example, in the form of $SU(5)$ fermionic singlets ($\bm{1}$) without the need to enlarge the scalar sector. Note that one can also introduce additional non-trivial representations of matter to generate observed fermion mass parameters in the lepton~\cite{Bajc:2006ia} and quark~\cite{Oshimo:2009ia} sectors. That, however, would not alter our operator analysis for large enough masses of extra matter fields.

Relevant decomposition of scalar representations to the SM gauge group, i.e., $SU(5) \to SU(3) \times SU(2) \times U(1)$, is given below~\cite{Slansky:1981yr}:
\begin{itemize}
\item $\bm{5}= (\bm{1},\bm{2},1/2)\oplus(\bm{3},\bm{1},-1/3)$;
\item $\bm{10}= (\bm{1},\bm{1},1)\oplus(\overline{\bm{3}},\bm{1},-2/3)\oplus(\bm{3},\bm{2},1/6)$;
\item $\bm{15}= (\bm{1},\bm{3},1)\oplus(\bm{3},\bm{2},1/6)\oplus(\bm{6},\bm{1},-2/3)$;
\item $\bm{45}= (\bm{8},\bm{2},1/2)\oplus
(\overline{\bm{6}},\bm{1}, -1/3) \oplus (\bm{3},\bm{3},-1/3)
\oplus (\overline{\bm{3}}, \bm{2}, -7/6) \oplus (\bm{3},\bm{1},
-1/3) \oplus (\overline{\bm{3}}, \bm{1}, 4/3) \oplus (\bm{1},
\bm{2}, 1/2)$;
\item $\bm{50}= (\bm{8},\bm{2},1/2)\oplus
(\bm{6},\bm{1}, 4/3) \oplus (\overline{\bm{6}},\bm{3},-1/3)
\oplus (\overline{\bm{3}}, \bm{2}, -7/6) \oplus (\bm{3}, \bm{1}, -1/3) \oplus (\bm{1},
\bm{1}, -2)$.
\end{itemize}

Only $\bm{5}$, $\bm{15}$ and $\bm{45}$ contain electrically neutral components and are thus capable 
of developing phenomenologically viable vacuum expectation values (VEVs). Contributions to the up-quark, down-quark and charged lepton masses can come from both $\bm{5}$ and
$\bm{45}$ whereas Majorana (Dirac) masses for neutrinos can be generated by VEV of $\bm{15}$ ($\bm{5}$).

The scalar leptoquark states that violate both $B$ and $L$ quantum
numbers are $(\bm{3},\bm{1},-1/3)$, $(\bm{3},\bm{3},-1/3)$ and
$(\overline{\bm{3}}, \bm{1}, 4/3)$, if one {\it assumes} neutrinos to
be Majorana particles. These states reside in $\bm{5}$, $\bm{45}$ and
$\bm{50}$. However, if one allows for the possibility that neutrinos
are Dirac particles there is another
leptoquark---$(\overline{\bm{3}},\bm{1},-2/3)$---that is found in the
$\bm{10}$ of $SU(5)$ that violates both $B$ and $L$ and could thus
also destabilize proton. To that end we consider both the Majorana and
Dirac neutrino cases to keep the analysis as general as
possible. Altogether, there are eighteen (fifteen) scalar leptoquarks that could mediate proton decay in case neutrinos are Dirac (Majorana) particles. The leptoquarks in question are all triplets of color as they must contract with lepton and quark states into an $SU(3)$ singlet. Tables~\ref{table:SU(5)-50},~\ref{table:SU(5)-45},~\ref{table:SU(5)-10}
and~\ref{table:SU(5)-5} summarize couplings to the matter of relevant
states that reside in $50$-, $45$-, $10$- and $5$-dimensional
representations, respectively.
\begin{table}[htdp]
\begin{center}
\begin{tabular}{| c || c |}\hline
 $SU(5)$   &  $Y^{10}_{ij} \bm{10}_{i} \bm{10}_{j}\bm{50}$\\\hline
  $(\bm{3},\bm{1},-1/3)$  & \\
   $\equiv$  & \raisebox{2.4ex}[0pt]{$12^{-1/2} \epsilon_{abc}
[Y^{10}_{ij} +Y^{10}_{ji}] d_{a\,i}^T C
 u_{b\,j} \Delta_{c}$}\\
   $\Delta$  &  \raisebox{2.4ex}[0pt]{$3^{-1/2} [Y^{10}_{ij} +Y^{10}_{ji}]
e^{C\,T}_{i} C u^C_{a\,j} \Delta_{a}$}\\\hline
\end{tabular}
\end{center}
\caption{Yukawa couplings of the $B$ and $L$ violating scalar in the $50$-dimensional representation of $SU(5)$. $a,b,c=1,2,3$ ($i,j=1,2,3$) are color (flavor) indices. $Y^{10}_{ij}$ are Yukawa matrix elements associated with the relevant contraction in the group space of $SU(5)$.}
\label{table:SU(5)-50}
\end{table}

\begin{table}[htdp]
\begin{center}
\begin{tabular}{| c || c | c |}\hline
$SU(5)$   &  $Y^{10}_{ij} \bm{10}_{i} \bm{10}_{j}\bm{45}$ & $Y^{\overline{5}}_{ij} \bm{10}_{i} \overline{\bm{5}}_{j}\bm{45}^*$\\\hline
$(\bm{3},\bm{1},-1/3)$  & & $2^{-1} Y^{\overline{5}}_{ij} \epsilon_{abc} u_{a\,i}^{C\,T} C  d^C_{b\,j} \Delta^{*}_{c}$\\
$\equiv$  &  $2^{1/2} [Y^{10}_{ij} -Y^{10}_{ji}] e^{C\,T}_{i} C u^C_{a\,j} \Delta_{a}$ &$-2^{-1} Y^{\overline{5}}_{ij} u_{a\,i}^T C e_{j} \Delta^{*}_{a}$\\
$\Delta$  & & $2^{-1} Y^{\overline{5}}_{ij} d_{a\,i}^T C \nu_{j}\Delta^{*}_{a}$\\\hline
               & $2^{1/2} \epsilon_{abc}
[Y^{10}_{ij}-Y^{10}_{ji}] d_{a \, i}^T C
d_{b\,j}\Delta^{1}_{c}$ & $Y^{\overline{5}}_{ij} u_{a\,i}^T C \nu_j \Delta^{1*}_{a}$\\\cline{2-3}
   \raisebox{2.4ex}[0pt]{$(\bm{3},\bm{3},-1/3)$}  &  & $2^{-1/2} Y^{\overline{5}}_{ij} u_{a \, i}^T C e_j \Delta^{2*}_{a}$
\\ \raisebox{2.4ex}[0pt]{$\equiv$}&  \raisebox{2.4ex}[0pt]{$- 2 \epsilon_{abc}
[Y^{10}_{ij}-Y^{10}_{ji}] d_{a \, i}^T C u_{b\,j} \Delta^{2}_{c}$} & $2^{-1/2}Y^{\overline{5}}_{ij} d_{a \, i}^T C \nu_j \Delta^{2*}_{a}$\\\cline{2-3}
\raisebox{2.4ex}[0pt]{$(\Delta^{1},\Delta^{2},\Delta^{3})$}& $-2^{1/2} \epsilon_{abc}
[Y^{10}_{ij}-Y^{10}_{ji}] u_{a \, i}^T C
u_{b\,j}\Delta^{3}_{c}$ & $- Y^{\overline{5}}_{ij} d_{a\,i}^T C e_j \Delta^{3*}_{a}$\\\hline
$(\overline{\bm{3}}, \bm{1}, 4/3)$  &  & \\
$\equiv$  &  $2^{1/2}
[Y^{10}_{ij} -Y^{10}_{ji}] \epsilon_{abc} u^{C\,T}_{i\,a} C u^C_{b\,j} \Delta_{c}$ & $
-Y^{\overline{5}}_{ij} e_{i}^{C\,T} C  d^C_{a\,j}
\Delta^{*}_{a}$\\
  $\Delta$  &  & \\\hline\end{tabular}
\end{center}
\caption{Yukawa couplings of the $B$ and $L$ violating scalars in the $45$-dimensional representation of $SU(5)$. $a,b,c=1,2,3$ ($i,j=1,2,3$) are color (flavor) indices. $Y^{10}_{ij}$ and $Y^{\overline{5}}_{ij}$ are Yukawa matrix elements.}
\label{table:SU(5)-45}
\end{table}

\begin{table}[htdp]
\begin{center}
\begin{tabular}{| c || c | c |}\hline
 $SU(5)$   &  $Y^{1}_{ij} \bm{10}_{i} \bm{1}_{j} \bm{10}^{*}$ & $Y^{\overline{5}}_{ij} \overline{\bm{5}}_{i} \overline{\bm{5}}_{j} \bm{10}$\\\hline
  $(\overline{\bm{3}},\bm{1},-2/3)$  & & \\
   $\equiv$  & $Y^{1}_{ij}
 u^{C\,T}_{a\,i} C
 \nu^C_{j} \Delta^*_{a}$ & $ 2^{-1/2}\epsilon_{abc} Y^{\overline{5}}_{ij}
 d^{C\,T}_{a\,i} C
 d^C_{b\,j} \Delta_{c}$\\
   $\Delta$  & & $Y^{\overline{5}}=-Y^{\overline{5}\,T}$\\\hline
\end{tabular}
\end{center}
\caption{Yukawa couplings of the $B$ and $L$ violating scalar in the $10$-dimensional representation of $SU(5)$. $a,b,c=1,2,3$ ($i,j=1,2,3$) are color (flavor) indices. $Y^{1}_{ij}$ and $Y^{\overline{5}}_{ij}$ are Yukawa matrix elements.}
\label{table:SU(5)-10}
\end{table}

\begin{table}[htdp]
\begin{center}
\begin{tabular}{| c || c | c | c |}\hline
 $SU(5)$   &  $Y^{10}_{ij} \bm{10}_{i} \bm{10}_{j}\bm{5}$ & $Y^{\overline{5}}_{ij} \bm{10}_{i} \overline{\bm{5}}_{j}\bm{5}^*$ & $Y^{1}_{ij} \overline{\bm{5}}_{i} \bm{1}_{j}\bm{5}$\\\hline
  $(\bm{3},\bm{1},-1/3)$  & & $ 2^{-1/2}\epsilon_{abc} Y^{\overline{5}}_{ij}
 u^{C\,T}_{a\,i} C
 d^C_{b\,j} \Delta^*_{c}$ &\\
   $\equiv$  & \raisebox{2.4ex}[0pt]{$2 \epsilon_{abc}
[Y^{10}_{ij} +Y^{10}_{ji}] d_{a\,i}^T C
 u_{b\,j} \Delta_{c}$} & $2^{-1/2}Y^{\overline{5}}_{ij} u_{a\,i}^T C e_{j} \Delta^*_{a}$ & $ Y^{1}_{ij}
 d^{C\,T}_{a\,i} C
 \nu^C_{j} \Delta_{a}$\\
   $\Delta$  & \raisebox{2.4ex}[0pt]{$-2 [Y^{10}_{ij} +Y^{10}_{ji}]
e^{C\,T}_{i} C u^C_{a\,j} \Delta_{a}$} & $-2^{-1/2}Y^{\overline{5}}_{ij} d_{a\,i}^T C \nu_{j} \Delta^*_{a}$ &\\\hline
\end{tabular}
\end{center}
\caption{Yukawa couplings of the $B$ and $L$ violating scalar in the $5$-dimensional representation of $SU(5)$. $a,b,c=1,2,3$ ($i,j=1,2,3$) are color (flavor) indices. $Y^{10}_{ij}$, $Y^{\overline{5}}_{ij}$ and $Y^{1}_{ij}$ are Yukawa matrix elements.}
\label{table:SU(5)-5}
\end{table}

We observe that in the $SU(5)$ framework the primary obstacle
to the proton stability seems to be the need to generate Yukawa
couplings relevant for the charged lepton and down-quark masses. These
receive equally important contributions from the $\bm{10}_{i}
\overline{\bm{5}}_{j}\bm{5}^*$ and $\bm{10}_{i}
\overline{\bm{5}}_{j}\bm{45}^*$ contractions~\cite{Georgi:1979df}. It is clear from
Tables~\ref{table:SU(5)-45} and~\ref{table:SU(5)-5} that both of
these, individually, generate potentially dangerous couplings. The
up-quark Yukawa coupling generation, on the other hand, via the
$\bm{10}_{i} \bm{10}_{j}\bm{45}$ operator seems not to pose any danger
whatsoever as can be seen from the second column in
Table~\ref{table:SU(5)-45}. However, that operator cannot generate
viable masses for all up-quarks due to the antisymmetry of the
corresponding mass matrix. The $\bm{10}_{i} \bm{10}_{j}\bm{5}$
contraction does provide viable up-quark masses but the price to pay
is resurrection of the proton decay issue. To conclude, the only
operator that can be considered innocuous in the Majorana neutrino case is
the $\bm{10}_{i} \bm{10}_{j}\bm{45}$ contraction.

\section{Leptoquarks in flipped $SU(5)$}
\label{flippedSU(5)}
Another possibility to unify
the SM matter into an $SU(5)$-based framework leads to the so-called flipped $SU(5)$ scenario~\cite{DeRujula:1980qc,Georgi:1980pw,Barr:1981qv}. A single family of matter fields in 
flipped $SU(5)$ can be seen as originating from a $16$-dimensional representation of $SO(10)$.
Actually, flipped $SU(5)$ is not necessarily completely embedded into $SO(10)$. Nevertheless, the generator of electric charge in flipped $SU(5)$ is given as a linear combination of a $U(1)$ generator that resides in $SU(5)$ and an extra $U(1)$ generator as if both of these originate from an $SO(10) \rightarrow SU(5) \times U(1)$ decomposition. This guarantees anomaly cancelation at the price of introducing one extra state per family, i.e., the right-handed neutrino $\nu^C$. The transition between the $SU(5)$ and flipped $SU(5)$ embeddings is then provided by $d^C \leftrightarrow u^C$, $e^C \leftrightarrow
\nu^C$, $u \leftrightarrow d$ and $\nu \leftrightarrow e$ transformations. Flipped $SU(5)$ thus predicts existence of three right-handed neutrinos as these transform nontrivially under the underlying gauge symmetry.

The matter fields in flipped $SU(5)$ comprise $\bm{10}_i^{+1}$,
$\overline{\bm{5}}_i^{-3}$ and $\bm{1}_i^{+5}$, where the superscripts
correspond to the extra $U(1)$ charge assignment. To obtain the SM
hypercharge $Y$ one uses the relation $Y=(Y(U(1))-Y(U(1)_{SU(5)}))/5$,
where $Y(U(1))$ and $Y(U(1)_{SU(5)})$ represent the quantum numbers of
the extra $U(1)$ and the $U(1)$ in $SU(5)(\rightarrow SU(3) \times
SU(2) \times U(1))$, respectively. One then obtains the electric charge
as usual, $Q = Y + T_3$.

The scalar sector that can couple to matter directly is made out of
$\bm{50}^{-2}$, $\bm{45}^{-2}$, $\bm{15}^{+6}$, $\bm{10}^{+6}$,
$\bm{5}^{-2}$ and $\bm{1}^{-10}$. Representations that can generate
contributions to the charged fermion masses and Dirac neutrino masses
are $\bm{45}^{-2}$ and $\bm{5}^{-2}$, whereas Majorana mass for
neutrinos can originate from interactions with
$\bm{15}^{+6}$. Leptoquarks that violate $B$ and $L$ reside in
$\bm{50}^{-2}$, $\bm{45}^{-2}$, $\bm{10}^{+6}$ and $\bm{5}^{-2}$ with
relevant couplings to matter given in
Tables~\ref{table:flippedSU(5)-50},~\ref{table:flippedSU(5)-45},~\ref{table:flippedSU(5)-5}
and~\ref{table:flippedSU(5)-10}, respectively.

\begin{table}[htdp]
\begin{center}
\begin{tabular}{| c || c |}\hline
 $SU(5) \times U(1)$   &  $Y^{10}_{ij} \bm{10}_{i}^{+1} \bm{10}_{j}^{+1} \bm{50}^{-2}$\\\hline
  $(\bm{3},\bm{1},-1/3)^{-2}$  & \\
   $\equiv$  &  \raisebox{2.4ex}[0pt]{$12^{-1/2} \epsilon_{abc}
[Y^{10}_{ij} +Y^{10}_{ji}] u_{a\,i}^T C
 d_{b\,j} \Delta_{c}$}\\
   $\Delta$  &  \raisebox{2.4ex}[0pt]{$3^{-1/2} [Y^{10}_{ij} +Y^{10}_{ji}]
\nu^{C\,T}_{i} C d^C_{a\,j} \Delta_{a}$}\\\hline
\end{tabular}
\end{center}
\caption{Yukawa couplings of the $B$ and $L$ violating scalar in $50$-dimensional representation of $SU(5)$. $a,b,c=1,2,3$ ($i,j=1,2,3$) are color (flavor) indices. $Y^{10}_{ij}$ are Yukawa matrix elements.}
\label{table:flippedSU(5)-50}
\end{table}

\begin{table}[htdp]
\begin{center}
\begin{tabular}{| c || c | c |}\hline
 $SU(5) \times U(1)$   &  $Y^{10}_{ij} \bm{10}_{i}^{+1} \bm{10}_{j}^{+1}\bm{45}^{-2}$ & $Y^{\overline{5}}_{ij} \bm{10}_{i} \overline{\bm{5}}_{j}^{-3}\bm{45}^{*\,+2}$\\\hline
  $(\bm{3},\bm{1},-1/3)^{-2}$  & & $2^{-1}
Y^{\overline{5}}_{ij} \epsilon_{abc} d_{a\,i}^{C\,T} C  u^C_{b\,j}
\Delta^{*}_{c}$\\
   $\equiv$  &  $2^{1/2}
[Y^{10}_{ij} -Y^{10}_{ji}] \nu^{C\,T}_{i} C d^C_{a\,j} \Delta_{a}$ & $-2^{-1} Y^{\overline{5}}_{ij} d_{a\,i}^T C
 \nu_{j} \Delta^{*}_{a}$\\
   $\Delta$  & & $2^{-1} Y^{\overline{5}}_{ij} u_{a\,i}^T C
 e_{j}\Delta^{*}_{a}$\\\hline
   & $2^{1/2} \epsilon_{abc}
[Y^{10}_{ij}-Y^{10}_{ji}] u_{a \, i}^T C
u_{b\,j}\Delta^{3}_{c}$ & $Y^{\overline{5}}_{ij} d_{a\,i}^T C e_j \Delta^{3*}_{a}$\\\cline{2-3}
   \raisebox{2.4ex}[0pt]{$(\bm{3},\bm{3},-1/3)^{-2}$}  &  & $2^{-1/2} Y^{\overline{5}}_{ij} d_{a \, i}^T C \nu_j \Delta^{2*}_{a}$\\
 \raisebox{2.4ex}[0pt]{$\equiv$}&  \raisebox{2.4ex}[0pt]{$-2 \epsilon_{abc}
[Y^{10}_{ij}-Y^{10}_{ji}] u_{a \, i}^T C d_{b\,j} \Delta^{2}_{c}$} & $2^{-1/2}Y^{\overline{5}}_{ij} u_{a \, i}^T C e_j \Delta^{2*}_{a}$\\\cline{2-3}
 \raisebox{2.4ex}[0pt]{$(\Delta^{1},\Delta^{2},\Delta^{3})$}& $-2^{1/2} \epsilon_{abc}
[Y^{10}_{ij}-Y^{10}_{ji}] d_{a \, i}^T C
d_{b\,j}\Delta^{1}_{c}$ & $- Y^{\overline{5}}_{ij} u_{a\,i}^T C \nu_j \Delta^{1*}_{a}$\\\hline
$(\overline{\bm{3}}, \bm{1}, 4/3)^{-2}$  & & \\
$\equiv$  &  $2^{1/2}[Y^{10}_{ij} -Y^{10}_{ji}] \epsilon_{abc} d^{C\,T}_{i\,a} C d^C_{b\,j} \Delta_{c}$ & $-Y^{\overline{5}}_{ij} \nu_{i}^{C\,T} C u^C_{a\,j}
\Delta^{*}_{a}$\\
  $\Delta$  &  & \\\hline\end{tabular}
\end{center}
\caption{Yukawa couplings of the $B$ and $L$ violating scalars in $45$-dimensional representation of $SU(5)$. $a,b,c=1,2,3$ ($i,j=1,2,3$) are color (flavor) indices. $Y^{10}_{ij}$ and $Y^{\overline{5}}_{ij}$ are Yukawa matrix elements.}
\label{table:flippedSU(5)-45}
\end{table}

\begin{table}[htdp]
\begin{center}
\begin{tabular}{| c || c | c |}\hline
 $SU(5) \times U(1)$   &  $Y^{1}_{ij} \bm{10}_{i}^{+1} \bm{1}_{j}^{+5}\bm{10}^{*\,-6}$ & $Y^{\overline{5}}_{ij} \overline{\bm{5}}_{i}^{-3} \overline{\bm{5}}_{j}^{-3}\bm{10}^{+6}$\\\hline
  $(\overline{\bm{3}},\bm{1},-2/3)^{+6}$  & & \\
   $\equiv$  & $Y^{1}_{ij}
 d^{C\,T}_{a\,i} C
 e^C_{j} \Delta^*_{a}$ & $ 2^{-1/2}\epsilon_{abc} Y^{\overline{5}}_{ij}
 u^{C\,T}_{a\,i} C
 u^C_{b\,j} \Delta_{c}$\\
   $\Delta$  & & $Y^{\overline{5}}=-Y^{\overline{5}\,T}$\\\hline
\end{tabular}
\end{center}
\caption{Yukawa couplings of the $B$ and $L$ violating scalar in $10$-dimensional representation of $SU(5)$. $a,b,c=1,2,3$ ($i,j=1,2,3$) are color (flavor) indices. $Y^{1}_{ij}$ and $Y^{\overline{5}}_{ij}$ are Yukawa matrix elements.}
\label{table:flippedSU(5)-10}
\end{table}

\begin{table}[htdp]
\begin{center}
\begin{tabular}{| c || c | c | c |}\hline
 $SU(5) \times U(1)$   &  $Y^{10}_{ij} \bm{10}_{i}^{+1} \bm{10}_{j}^{+1}\bm{5}^{-2}$ & $Y^{\overline{5}}_{ij} \bm{10}_{i}^{+1} \overline{\bm{5}}_{j}^{-3}\bm{5}^{*\,+2}$& $Y^{1}_{ij} \overline{\bm{5}}_{i}^{-3}\bm{1}_{j}^{+5} \bm{5}^{-2}$\\\hline
  $(\bm{3},\bm{1},-1/3)^{-2}$  & & $ 2^{-1/2}\epsilon_{abc} Y^{\overline{5}}_{ij}
 d^{C\,T}_{a\,i} C
 u^C_{b\,j} \Delta^*_{c}$ &\\
   $\equiv$  & \raisebox{2.4ex}[0pt]{$-2 \epsilon_{abc}
[Y^{10}_{ij} +Y^{10}_{ji}] u_{a\,i}^T C
 d_{b\,j} \Delta_{c}$} & $2^{-1/2}Y^{\overline{5}}_{ij} d_{a\,i}^T C \nu_{j} \Delta^*_{a}$ & $ Y^{1}_{ij}
 u^{C\,T}_{a\,i} C
 e^C_{j} \Delta_{a}$\\
   $\Delta$  & \raisebox{2.4ex}[0pt]{$-2 [Y^{10}_{ij} +Y^{10}_{ji}]
\nu^{C\,T}_{i} C d^C_{a\,j} \Delta_{a}$} & $-2^{-1/2}Y^{\overline{5}}_{ij} u_{a\,i}^T C e_{j} \Delta^*_{a}$ &\\\hline
\end{tabular}
\end{center}
\caption{Yukawa couplings of the $B$ and $L$ violating scalar in $5$-dimensional representation of $SU(5)$. $a,b,c=1,2,3$ ($i,j=1,2,3$) are color (flavor) indices. $Y^{10}_{ij}$, $Y^{\overline{5}}_{ij}$ and $Y^{1}_{ij}$ are Yukawa matrix elements.}
\label{table:flippedSU(5)-5}
\end{table}

In flipped $SU(5)$ the main obstacle to matter stability is the the
generation of the up-quark masses. Namely, these can be generated through $\bm{10}_{i}^{+1} \overline{\bm{5}}_{j}^{-3}\bm{5}^{*\,+2}$ and/or
$\bm{10}_{i}^{+1} \overline{\bm{5}}_{j}^{-3}\bm{45}^{*\,+2}$
contractions. Both of these are dangerous as far as the proton decay
is concerned as can be seen from Tables~\ref{table:flippedSU(5)-45}
and~\ref{table:flippedSU(5)-5}. All other contractions, in the Majorana
neutrino case, are actually innocuous.

\section{Proton decay}
\label{proton_decay}
Let us discuss proton decay operators due to the scalar leptoquark
exchange of the lowest possible dimension in detail. These are
dimension-six operators made out of three quarks and a lepton that
violate $B$ and $L$ by 1 unit. They are summarized below
\begin{eqnarray}
\textit{O}_H (d_{\alpha}, e_{\beta})& = & a(d_{\alpha}, e_{\beta}) \ u^T \ L \ C^{-1} \ d_{\alpha} \ u^T \ L \ C^{-1} e_{\beta}\,, \\
\textit{O}_H (d_{\alpha}, e_{\beta}^C)& = & a(d_{\alpha}, e_{\beta}^C)\ u^T \ L \ C^{-1} \ d_{\alpha} \ 
{e^C_{\beta}}^{\dagger} \ L \ C^{-1} {u^C}^* \,,\\
\textit{O}_H (d_{\alpha}^C, e_{\beta})& = & a(d_{\alpha}^C, e_{\beta}) \ {d^C_{\alpha}}^{\dagger} \ L \ C^{-1} \ {u^C}^* 
\ u^T \ L \ C^{-1} e_{\beta} \,,\\
\textit{O}_H (d_{\alpha}^C, e_{\beta}^C)& = & a(d_{\alpha}^C, e_{\beta}^C) \ {d^C_{\alpha}}^{\dagger} \ L \ C^{-1} \ {u^C}^* 
\ {e^C_{\beta}}^{\dagger} \ L \ C^{-1} {u^C}^*\,,\\
\textit{O}_H (d_{\alpha}, d_{\beta}, \nu_i) & = & a(d_{\alpha}, d_{\beta}, \nu_i) \ u^T \ L \ C^{-1} \ d_{\alpha} \ d_{\beta}^T \ L \ C^{-1}\ \nu_i \,,\\
\textit{O}_H (d_{\alpha}, d_{\beta}^C, \nu_i) & = & a(d_{\alpha}, d_{\beta}^C, \nu_i) \ {d^C_{\beta}}^{\dagger} \ L \ C^{-1} \ {u^C}^* 
\ d_{\alpha}^T \ L \ C^{-1}\ \nu_i \,,\\
\textit{O}_H (d_{\alpha}, d^C_{\beta}, \nu^C_i) & = & a(d_{\alpha}, d_{\beta}^C, \nu^C_i) \ u^T \ L \ C^{-1} \ d_{\alpha} \ {\nu^C_{i}}^{\dagger} \ L \ C^{-1}\ {d_{\beta}^C}^* \,,\\
\textit{O}_H (d^C_{\alpha}, d_{\beta}^C, \nu^C_i) & = & a(d_{\alpha}^C, d_{\beta}^C, \nu^C_i) \ {d^C_{\beta}}^{\dagger} \ L \ C^{-1} \ {u^C}^* 
\ {\nu^C_{i}}^{\dagger}  \ L \ C^{-1}\ {d_{\alpha}^C}^*  \,.
\end{eqnarray}
Here, $i(=1,2,3)$ and $\alpha,\beta(=1,2)$ are generation indices,
where all operators that involve a neutrino are bound to have
$\alpha+\beta<4$ due to kinematical constraints. $L(=(1- \gamma_5)/2)$
is the left projection operator.  The $SU(3)$ color indices are not
shown since the antisymmetric contraction $\epsilon_{abc} q_a q_b q_c$
is common to all the above listed operators. This notation has already
been introduced in Ref.~\cite{Nath:2006ut}.

These operators allow one to write down explicitly $d=6$ proton decay contributions due to a particular leptoquark exchange~\cite{Nath:2006ut}. To that end we first specify our convention for the redefinition of the fermion fields that yields the up-, down-quark and charged lepton mass matrices in physical basis: $M_{U,D,E} \rightarrow M^{\textrm{diag}}_{U,D,E}$. These are $U^T_C M_U U = M_U^{\textrm{diag}}$, $ D^T_C M_D D= M_D^{\textrm{diag}}$, and $E^T_C M_E E = M_E^{\textrm{diag}}$.
The quark mixing is $U^{\dagger} D \equiv V_{UD}=K_1 V_{CKM}
K_2$, where $K_1$ and $K_2$ are diagonal matrices containing three
and two phases, respectively. In the neutrino sector we have $N^T_C
M_N N = M_N^{\textrm{diag}}$  ($N^T M_N N = M_N^{\textrm{diag}}$) in
the case of Dirac (Majorana) neutrinos. The leptonic mixing $E^\dagger
N  \equiv V_{EN}= K_3
V_{PMNS} K_4$ in case of Dirac neutrino, or $V_{EN}=K_3 V_{PMNS}$ in the
Majorana case. $K_3$ is a diagonal matrix containing three
phases whereas $K_4$ contains two phases. $V_{CKM}$ ($V_{PMNS}$) is the Cabibbo-Kobayashi-Maskawa (Pontecorvo-Maki-Nakagawa-Sakata) mixing
matrix. 

\subsection{Tree-level exchange ($d=6$) operators in $SU(5)$}
\label{SU(5)_OPERATORS}
The only relevant coefficient for $\Delta \equiv (\bm{3},\bm{1},-1/3)$ from $\bm{50}$ is
\begin{eqnarray}
a(d_{\alpha}, e_{\beta}^C) &=& \frac{1}{6 m_\Delta^2} \ (U^T (Y^{10}+ Y^{10\,T}) D)_{1 \alpha} 
\ (E_C^{\dagger} (Y^{10}+ Y^{10\,T})^{\dagger} U_C^*)_{\beta 1},
\end{eqnarray}
where $m_\Delta$ is a mass of leptoquark in question. (See Table~\ref{table:SU(5)-50} for details on notation for Yukawa couplings of the $50$-dimensional representation to the matter.)

The relevant coefficients for $\Delta \equiv (\bm{3},\bm{1},-1/3)$ from $\bm{45}$ are
\begin{eqnarray}
a(d_{\alpha}^C, e_{\beta}) &=& \frac{1}{4 m_\Delta^2} \ (D_C^{\dagger} Y^{\overline{5}\,\dagger} U_C^*)_{\alpha 1} 
\ (U^T Y^{\overline{5}} E)_{1 \beta}\,,\\
a(d_{\alpha}^C, e_{\beta}^C) &=& \frac{1}{\sqrt{2} m_\Delta^2} \ (D_C^{\dagger} Y^{\overline{5}\,\dagger} U_C^*)_{\alpha 1} 
\ (E_C^{\dagger} (Y^{10}- Y^{10\,T})^{\dagger} U_C^*)_{\beta 1}\,,\\
a (d_{\alpha}, d_{\beta}^C, \nu_i) &=&\frac{1}{4 m_\Delta^2} \ (D_C^{\dagger} Y^{\overline{5}\,\dagger} U_C^*)_{\beta 1} \ 
(D^T Y^{\overline{5}} N)_{\alpha i}\,.
\end{eqnarray}

The relevant coefficients for $\Delta \equiv (\bm{3},\bm{1},-1/3)$ from $\bm{5}$ are
\begin{eqnarray}
\label{T_1}
a(d_{\alpha}, e_{\beta}) &=& -\frac{\sqrt{2}}{m_\Delta^2} \ (U^T ( Y^{10}+Y^{10\,T}) D)_{1 \alpha} 
\ (U^T Y^{\overline{5}} E)_{1 \beta}\,,\\
\label{T_2}
a(d_{\alpha}, e_{\beta}^C) &=& -\frac{4}{m_\Delta^2} \ (U^T (Y^{10}+Y^{10\,T}) D)_{1 \alpha} 
\ (E_C^{\dagger} (Y^{10}+Y^{10\,T})^{\dagger} U_C^*)_{\beta 1}\,,\\
\label{T_3}
a(d_{\alpha}^C, e_{\beta}) &=& \frac{1}{2 m_\Delta^2} \ (D_C^{\dagger} Y^{\overline{5}\,\dagger} U_C^*)_{\alpha 1} 
\ (U^T Y^{\overline{5}} E)_{1 \beta}\,,\\
\label{T_4}
a(d_{\alpha}^C, e_{\beta}^C) &=& \frac{\sqrt{2}}{m_\Delta^2} \ (D_C^{\dagger} Y^{\overline{5}\,\dagger} U_C^*)_{\alpha 1} 
\ (E_C^{\dagger} (Y^{10}+Y^{10\,T})^{\dagger} U_C^*)_{\beta 1}\,,\\
\label{T_5}
a(d_{\alpha}, d_{\beta}, \nu_i)&=& \frac{\sqrt{2}}{m_\Delta^2} (U^T ( Y^{10}+Y^{10\,T}) D)_{1 \alpha} 
\ (D^T Y^{\overline{5}} N)_{\beta i}\,,\\
\label{T_6}
a (d_{\alpha}, d_{\beta}^C, \nu_i) &=& -\frac{1}{2 m_\Delta^2} \ (D_C^{\dagger} Y^{\overline{5}\,\dagger} U_C^*)_{\beta 1} \ 
(D^T Y^{\overline{5}} N)_{\alpha i}\,,\\
\label{T_7}
a(d_{\alpha}, d_{\beta}^C, \nu^C_i)&=&\frac{2}{m_\Delta^2} (U^T ( Y^{10}+Y^{10\,T}) D)_{1 \alpha} 
\ (N_C^{\dagger} Y^{1\,\dagger} D^*_C)_{i \beta}\,,\\
\label{T_8}
a(d_{\alpha}^C, d_{\beta}^C, \nu^C_i)&=&-\frac{1}{\sqrt{2} m_\Delta^2} (D_C^{\dagger} Y^{\overline{5}\,\dagger} U_C^{*})_{\beta 1} 
\ (N_C^{\dagger} Y^{1\,\dagger} D_C^{*})_{i \alpha}\,.
\end{eqnarray}

The relevant coefficients for $\Delta^2 \in (\bm{3},\bm{3},-1/3)$ from $\bm{45}$ are
\begin{eqnarray}
a(d_{\alpha}, e_{\beta}) &=& -\frac{\sqrt{2}}{M_{\Delta^2}^2} \ (U^T (Y^{10}-Y^{10\,T}) D)_{1 \alpha} 
\ (U^T Y^{\overline{5}} E)_{1 \beta}\,,\\
a(d_{\alpha}, d_{\beta}, \nu_i)&=& -\frac{\sqrt{2}}{M_{\Delta^2}^2} (U^T ( Y^{10} - Y^{10\,T}) D)_{1 \alpha} 
\ (D^T Y^{\overline{5}} N)_{\beta i}\,.
\end{eqnarray}

The relevant coefficient for $\Delta^1 \in (\bm{3},\bm{3},-1/3)$ from $\bm{45}$ is
\begin{eqnarray}
a(d_{\alpha}, d_{\beta}, \nu_i)&=& \frac{2 \sqrt{2}}{M_{\Delta^1}^2} (U^T Y^{\overline{5}} N)_{1 i} 
\ (D^T ( Y^{10} - Y^{10\,T}) D)_{\beta \alpha}\,,
\end{eqnarray}
where the extra factor of $2$ comes from two terms in Fierz
transformation
\begin{equation}
 \label{eq:Fierz}
 (\overline{s^C} L d) (\overline{\nu^C} L u) = -
 (\overline{u^C} L s) (\overline{\nu^C} L d) - (\overline{u^C} L d) (\overline{\nu^C} L s)\,. 
\end{equation}

The only relevant coefficient for $\Delta \equiv (\overline{\bm{3}},\bm{1},-2/3)$ from $\bm{10}$ is
\begin{eqnarray}
a(d^C_{\alpha}, d^C_{\beta}, \nu_{i}^C) &=&  -\frac{1}{\sqrt{2} m_\Delta^2} \ (D_C^{\dagger} (Y^{\overline{5}}-Y^{\overline{5}\,T})^{\dagger} D_C^*)_{\beta \alpha} \ 
(N_C^{\dagger} Y^{1\,\dagger} U_C^*)_{i 1}\,.
\end{eqnarray}

Finally, $\Delta^3 \in (\bm{3},\bm{3},-1/3)$ and
$(\overline{\bm{3}},\bm{1},4/3)$, both from $\bm{45}$ of $SU(5)$, do not
contribute to proton decay at tree-level. This is due to
antisymmetry, in flavor space, of their couplings to the pair of up-quarks.
Nevertheless, both states still induce proton decay through loops at
an effective $d=6$ level. We present systematic study of these
contributions for the $(\overline{\bm{3}},\bm{1},4/3)$ case in
Section~\ref{Higher_Order}. There we also spell out contributions of
the $(\overline{\bm{3}},\bm{1},4/3)$ leptoquark to dimension-nine
tree-level proton decay amplitudes. Equivalent contributions of
$\Delta^3 \in (\bm{3},\bm{3},-1/3)$
are not pursued since the components $\Delta^1$ and $\Delta^2$ from
the same state already contribute at leading order. In this manner,
higher order contributions of $\Delta^{1,2,3}$ would only play a role
of radiative corrections.

\subsection{Tree-level exchange ($d=6$) operators in flipped $SU(5)$}
The only relevant coefficient for $\Delta \equiv (\bm{3},\bm{1},-1/3)^{-2}$ from $\bm{50}^{-2}$ is
\begin{eqnarray}
a(d_{\alpha}, d^C_{\beta}, \nu_{i}^C) &=& \frac{1}{6 m_\Delta^2} \ (U^T (Y^{10}+ Y^{10\,T}) D)_{1 \alpha} 
\ (N_C^{\dagger} (Y^{10}+ Y^{10\,T})^{\dagger} D_C^*)_{i \beta}\,.
\end{eqnarray}

The relevant coefficients for $\Delta \equiv (\bm{3},\bm{1},-1/3)^{-2}$ from $\bm{45}^{-2}$ are
\begin{eqnarray}
a(d_{\alpha}^C, e_{\beta}) &=& \frac{1}{4 m_\Delta^2} \ (D_C^{\dagger} Y^{\overline{5}\,*} U_C^*)_{\alpha 1} 
\ (U^T Y^{\overline{5}} E)_{1 \beta}\,,\\
a (d_{\alpha}, d_{\beta}^C, \nu_i) &=&-\frac{1}{4 m_\Delta^2} \ (D_C^{\dagger} Y^{\overline{5}\,*} U_C^*)_{\beta 1} \ 
(D^T Y^{\overline{5}} N)_{\alpha i}\,,\\
a(d_{\alpha}^C, d_{\beta}^C, \nu^C_i)&=&\frac{1}{\sqrt{2}m_\Delta^2} (D_C^* Y^{\overline{5}\,*} U_C^*)_{\beta 1} 
\ (N_C^\dagger (Y^{10}-Y^{10\,T})^\dagger D_C^*)_{i \alpha}\,.
\end{eqnarray}

The relevant coefficients for $\Delta \equiv (\bm{3},\bm{1},-1/3)^{-2}$ from $\bm{5}^{-2}$ are
\begin{eqnarray}
\label{F_1}
a(d_{\alpha}, e_{\beta}) &=& \frac{\sqrt{2}}{m_\Delta^2} \ (U^T ( Y^{10}+Y^{10\,T}) D)_{1 \alpha} 
\ (U^T Y^{\overline{5}} E)_{1 \beta}\,,\\
\label{F_2}
a(d_{\alpha}, e_{\beta}^C) &=& \frac{2}{m_\Delta^2} \ (U^T (Y^{10}+Y^{10\,T}) D)_{1 \alpha} 
\ (E_C^{\dagger} Y^{1\,\dagger} U_C^*)_{\beta 1}\,,\\
\label{F_3}
a(d_{\alpha}^C, e_{\beta}) &=& \frac{1}{2 m_\Delta^2} \ (D_C^{\dagger} Y^{\overline{5}\,*} U_C^*)_{\alpha 1} 
\ (U^T Y^{\overline{5}} E)_{1 \beta}\,,\\
\label{F_4}
a(d_{\alpha}^C, e_{\beta}^C) &=& \frac{1}{\sqrt{2} m_\Delta^2} \ (D_C^{\dagger} Y^{\overline{5}\,*} U_C^*)_{\alpha 1} 
\ (E_C^{\dagger} Y^{1\, \dagger} U_C^*)_{\beta 1}\,,\\
\label{F_5}
a (d_{\alpha}, d_{\beta}, \nu_i) &=&\frac{-\sqrt{2}}{m_\Delta^2} (U^T
(Y^{10}+Y^{10\,T}) D)_{1\alpha} \ (N^T Y^{\bar{5}\,T} D)_{i\beta}\,,\\
\label{F_6}
a (d_{\alpha}, d_{\beta}^C, \nu_i) &=&\frac{-1}{2 m_\Delta^2} \ (D_C^{\dagger} Y^{\overline{5}\,*} U_C^*)_{\beta 1} \ 
(D^T Y^{\overline{5}} N)_{\alpha i}\,,\\
\label{F_7}
a (d_{\alpha}, d_{\beta}^C, \nu_i^C) &=&\frac{-4}{m_\Delta^2} \ (U^T ( Y^{10}+Y^{10\,T}) D)_{1\alpha} \ 
(N_C^\dagger (Y^{10}+Y^{10\,T})^\dagger D_C^*)_{i\beta}\,,\\
\label{F_8}
a(d_{\alpha}^C, d_{\beta}^C, \nu^C_i)&=&-\frac{\sqrt{2}}{m_\Delta^2}
(D_C^\dagger Y^{\overline{5}\,*} U_C^*)_{1 \beta} 
\ (N_C^\dagger (Y^{10}+Y^{10\,T})^\dagger D_C^*)_{i \alpha}\,.
\end{eqnarray}

The relevant coefficients for $\Delta^2 \in (\bm{3},\bm{3},-1/3)^{-2}$ from $\bm{45}^{-2}$ are
\begin{eqnarray}
a(d_{\alpha}, e_{\beta}) &=& -\frac{\sqrt{2}}{M_{\Delta^2}^2} \ (U^T (Y^{10}-Y^{10\,T}) D)_{1 \alpha} 
\ (U^T Y^{\overline{5}} E)_{1 \beta}\,,\\
a(d_{\alpha}, d_{\beta}, \nu_i)&=& -\frac{\sqrt{2}}{M_{\Delta^2}^2} (U^T ( Y^{10} - Y^{10\,T}) D)_{1 \alpha} 
\ (D^T Y^{\overline{5}} N)_{\beta i}\,.
\end{eqnarray}

The relevant coefficient for $\Delta^1 \in (\bm{3},\bm{3},-1/3)^{-2}$ from $\bm{45}^{-2}$ is
\begin{eqnarray}
a(d_{\alpha}, d_{\beta}, \nu_i)&=& \frac{2 \sqrt{2}}{M_{\Delta^1}^2} (U^T Y^{\overline{5}} N)_{1 i} 
\ (D^T ( Y^{10} - Y^{10\,T}) D)_{\beta \alpha}\,,
\end{eqnarray}
where the extra factor of  $2$ comes from Fierz transformation~\eqref{eq:Fierz}.

The only relevant coefficient for $\Delta \equiv (\bm{3},\bm{1},4/3)^{-2}$ from $\bm{45}^{-2}$ is
\begin{eqnarray}
a(d^C_{\alpha}, d^C_{\beta}, \nu^C_i)&=& -\frac{2 \sqrt{2}}{M_{\Delta^1}^2} (U_C^\dagger Y^{\overline{5}\,\dagger} N_C^*)_{i1} 
\ (D_C^\dagger ( Y^{10} - Y^{10\,T})^\dagger D_C^*)_{\alpha \beta}\,,
\end{eqnarray}
where the extra factor of $2$ again comes from Fierz transformation.

The relevant coefficients for $(\overline{\bm{3}},\bm{1},-2/3)^{+6}
\in \bm{10}^{+6}$ and $\Delta^3 \in (\bm{3},\bm{3},-1/3)^{-2} \in
\bm{45}^{-2}$ are not present at the leading order due to antisymmetry
of the couplings to the up-quark pair. The higher order contributions
of the former state are discussed in Section~\ref{Higher_Order}.

\section{Leading order contributions}
\label{Discussion}
Scalar fields couple to matter through Yukawa couplings. This introduces uncertainties to predictions related to any process that involves scalar leptoquark exchange. It is thus natural to ask if, and when, it is necessary to have leptoquark masses close to a scale of grand unification. We will address this issue in the $SU(5)$ and flipped $SU(5)$ frameworks in what follows.
\subsection{Color triplets of $SU(5)$}
If the Yukawa sector relevant for proton decay through scalar exchange is not related to the origin of fermion masses and mixing parameters one cannot make any firm predictions. For example, all 
operators that correspond to the exchange of the triplet scalar in the $5$-dimensional representation of $SU(5)$ can be completely suppressed if $(U^T (Y^{10}+Y^{10\,T}) D)_{1 \alpha}=0$ and $(D_C^{\dagger}
Y^{\overline{5}\,\dagger} U_C^*)_{\alpha 1}=0$, $\alpha=1,2$, in the Majorana neutrino case. (Recall, it was the exchange of this scalar that has led to the so-called doublet-triplet splitting problem within the context of the Georgi-Glashow $SU(5)$ model~\cite{Georgi:1974sy}.) The suppression is certainly viable if the entries of $U$, $U_C$, $D$,
$D_C$, $Y^{10}$ and $Y^{\overline{5}}$ are all free parameters. The
first set of conditions can be insured if, for example,
$Y^{10}=-Y^{10\,T}$. This solution has already been pointed out in
Ref.~\cite{Nath:2006ut}. The second set of conditions can also be
easily satisfied although what we find defers from what has been
presented in~\cite{Nath:2006ut}.  

Things, however, change in models where the connection between Yukawa sector and fermion masses is strong. Let us thus analyze predictions of the simplest of all possible renormalizable models based on the $SU(5)$ gauge symmetry. We want to find what the current experimental bounds on the partial proton lifetimes for processes presented in Table~\ref{table:Proton_Decay_Bounds} imply for the masses of color triplets if the theory is to be viable with regard to the fermion mass generation. We analyze all these decay modes to make our study as complete as possible.
\begin{table}[h]
\begin{center}
\begin{tabular}{|l||r|}
\hline PROCESS & $\tau_p$ ($10^{33}$\,years)\\
\hline 
$p \rightarrow \pi^0 e^+ $ & $13.0$~\cite{Miura:2010zz}\\
$p \rightarrow \pi^0 \mu^+ $ & $11.0$~\cite{SuperKNew}\\
$p \rightarrow K^0 e^+$ & $1.0$~\cite{Kobayashi:2005pe}\\
$p \rightarrow K^0 \mu^+$ & $1.3$~\cite{Kobayashi:2005pe}\\
$p \rightarrow \eta e^+$ & $4.2$~\cite{Nishino:2012rv}\\
$p \rightarrow \eta \mu^+$ & $1.3$~\cite{Nishino:2012rv}\\
$p \rightarrow \pi^+  \bar{\nu}$ & $0.025$~\cite{Nakamura:2010zzi}\\
$p \rightarrow K^+ \bar{\nu}$ & $4.0$~\cite{Miura:2010zz}\\
\hline
\end{tabular}
\end{center}
\caption{Experimental bounds on selected partial proton decay lifetimes @ 90\,\% CL.}
\label{table:Proton_Decay_Bounds}
\end{table}

We demand in what follows that the theory is renormalizable and thus neglect possibility that higher-dimensional terms contribute to (super)potential at any level. We furthermore take the simplest possibility for the generation of phenomenologically viable fermion masses and mixing parameters. Namely, we demand that both $\bm{5}$ and $\bm{45}$ of Higgs contribute to the down-quark and charged lepton masses~\cite{Georgi:1979df}. We further take all mass matrices to be symmetric, i.e., $M_{U,D,E}=M_{U,D,E}^T$. This then allows us to consider two particular scenarios. The first (second) one represents the case when the contributions of the $(\bm{3},\bm{1},-1/3)$ state from the $5$-dimensional ($45$-dimensional) representation dominates. Our analysis is self-consistent as the symmetric mass matrix assumption eliminates contributions to proton decay of all other color triplets. Note also that any mixing between the triplets can be accounted for by simple rescaling of relevant operators.
\subsubsection{The charged anti-lepton final state}
We start our analysis with proton decay due to exchange of the triplet state from the $5$-dimensional representation. To find widths for the charged anti-leptons in the final state one needs to determine $a(d_{\alpha}, e_{\beta})$, $a(d_{\alpha}, e_{\beta}^C)$, $a(d_{\alpha}^C, e_{\beta})$ and  $a(d_{\alpha}^C, e_{\beta}^C)$. If the Yukawa couplings are symmetric the relevant input for these coefficients reads 
\begin{eqnarray}
\label{V_1}
(U^T ( Y^{10}+Y^{10\,T}) D)_{1 \alpha} &=& -\frac{1}{\sqrt{2} v_5} \ (M_U^{\textrm{diag}}V_{UD})_{1 \alpha}\,,\\
\label{V_2}
(U^T Y^{\overline{5}} E)_{1 \beta} &=& -\frac{1}{2 v_5} \ (3V^*_{UD} M_D^{\textrm{diag}} V^\dagger_{UD} U_2^*+U_2 M_E^{\textrm{diag}})_{1 \beta}\,,\\
\label{V_3}
(D^{\dagger} Y^{\overline{5}\,\dagger} U^*)_{\alpha 1} &=& -\frac{1}{2 v_5} \ (3 M_D^{\textrm{diag}} V^T_{UD}+V^\dagger_{UD}U^*_2 M_E^{\textrm{diag}} U^\dagger_2)_{\alpha 1}\,,\\
\label{V_4}
(E^{\dagger} (Y^{10}+Y^{10\,T})^{\dagger} U^*)_{\beta 1} &=&  -\frac{1}{\sqrt{2} v_5}\ (U_2^T M_U^{\textrm{diag}})_{\beta 1}\,,
\end{eqnarray}
where $U_2=U^T E^*$ and $v_5$ represents the VEV of the $5$-dimensional representation. $U_2$ entries and $v_5$ are primary sources of uncertainty. Our normalization is such that $|v_5|^2/2+12 |v_{45}|^2=v^2$, where $v(=246$\,GeV) stands for the electroweak VEV. $v_{45}$ is the VEV in the $45$-dimensional representation. A connection between Yukawa couplings and charged fermion mass matrices is spelled out elsewhere~\cite{Dorsner:2011ai}.

We outline details of our calculation using the $p \rightarrow e_\delta^+
\pi^0$ channels. Here, $\delta=1$ ($\delta=2$) corresponds to $e^+$ ($\mu^+$) in the final state. The decay widths formulas we use are summarized in Ref.~\cite{Nath:2006ut}. For these particular channels we have 
\begin{equation}
\Gamma(p \rightarrow e_\delta^+ \pi^0) = \frac{(m^2_p-m^2_{\pi^0})^2}{64 \pi f_\pi^2 m_p^3}\left(|\alpha a(d_1,e_\delta)+\beta a(d_1^C,e_\delta) |^2+|\alpha a(d_1,e_\delta^C)+\beta a(d_1^C,e_\delta^C) |^2\right)(1+D+F)^2,
\end{equation}
where $\alpha$ and $\beta$ are the so-called nucleon matrix elements. (See Section~\ref{Higher_Order} for more details on $\alpha$ and $\beta$.) $F + D$ and $F-D$ combinations are extracted from the nucleon axial charge and form factors in semileptonic hyperon decays, respectively~\cite{Claudson:1981gh, Aoki:2008ku}. We take $f_\pi =130$\,MeV, $m_p = 938.3$\,MeV, $D = 0.80(1)$, $F = 0.47(1)$ and $\alpha=-\beta =-0.0112(25)$\,GeV$^3$~\cite{Aoki:2008ku}. 

The uncertainty in predicting partial decay rates persists even in the minimal $SU(5)$ scenario with symmetric Yukawa couplings. This is evident from the $U_2$ dependence of $p \rightarrow e_\delta^+ \pi^0$ partial decay widths 
\begin{eqnarray*}
\Gamma(p \rightarrow e_\delta^+ \pi^0) &=& \frac{(m^2_p-m^2_{\pi^0})^2}{64 \pi f_\pi^2 m_p^3} \frac{\alpha^2}{v_5^4 m_\Delta^4}\left|(V_{UD})_{11}[m_u+\frac{3}{4}m_d] +\frac{1}{4} (V^\dagger_{UD} U_2^* M_E^{diag} U_2^\dagger)_{11}\right|^2\\&& \left(\left|\frac{3}{2}(V^*_{UD} M_D^{diag} V^\dagger_{UD} U_2^*)_{1 \delta}+\frac{1}{2} (U_2 M_E^{diag})_{1 \delta}\right|^2+4|m_u(U_2)_{1\delta}|^2\right) (1+D+F)^2.
\end{eqnarray*}
One can suppress (enhance) $\Gamma(p \rightarrow e_\delta^+ \pi^0)$ with regard to $U_2$ numerically to generate the least (most) conservative lower bound on the mass of the scalar triplet in the $5$-dimensional representation of $SU(5)$. This, however, should be done simultaneously with all other partial decay modes to generate a self-consistent solution. We will do that after we address proton decay into neutral anti-leptons in the final state. The important point is that even in the case of symmetric Yukawa couplings one cannot test the $SU(5)$ theory when the scalar triplet exchange dominates. This is in stark contrast to what one obtains for the gauge $d=6$ contributions in the $SU(5)$ framework~\cite{FileviezPerez:2004hn}.  

Nevertheless, we can already outline how one can find a maximum of $\Gamma(p \rightarrow e_\delta^+ \pi^0)$ with regard to $U_2$ to obtain the most conservative bound, from the model building point of view, on $m_\Delta$ without resorting to elaborate numerical analysis. The idea is to have Yukawa couplings of the third generation contribute as much as possible towards relevant amplitudes. The only possibility to achieve that is to make the 11 element of $U^*_2 M_E^{\textrm{diag}} U^\dagger_2$ matrix in Eq.~\eqref{V_3} as large as possible. This can be done with a simple ansatz 
\begin{equation}
\label{U2}
U_2 = U^T E^*= \left(\begin{array}{ccc}
0 & 0 & 1\\
0 & 1 & 0\\
1 & 0 & 0
\end{array}
\right),
\end{equation}
to obtain the following simplified expressions
\begin{eqnarray}
\label{A_1}
\Gamma(p \rightarrow e^+ \pi^0)&\approx&\frac{3^2 m_p \alpha^2}{2^{12} \pi f_\pi^2v_5^4 m_\Delta^4} 
\left|(V_{UD})_{13}\right|^2 m_b^2 m_\tau^2 (1+D+F)^2\,,\\
\label{A_2}
\Gamma(p \rightarrow \mu^+ \pi^0)&\approx&\frac{3^2 m_p \alpha^2}{2^{12} \pi f_\pi^2v_5^4 m_\Delta^4} 
\left|(V_{UD})_{12}\right|^2 m_s^2 m_\tau^2 (1+D+F)^2\,.
\end{eqnarray}
Interestingly enough, these expressions when combined with experimental input yield comparable bounds on the mass of the leptoquark in question. 

In the previous analysis it was assumed that contributions to proton decay of the triplet in the $5$-dimensional representation dominate over contributions of triplets in the $45$-dimensional representation. Let us now see if and when that is truly the case. Our assumption that the mass matrices are symmetric implies that the only other contribution to $p \rightarrow e_\delta^+ \pi^0$ ($\delta=1,2$) channels originates from exchange of the $(\bm{3},\bm{1},-1/3)$ state in the $45$-dimensional representation. In fact, the only relevant coefficient is $a(d_{\alpha}^C, e_{\beta})$ with the following entries 
\begin{eqnarray}
(D^{\dagger} Y^{\overline{5}\,\dagger} U^*)_{\alpha 1} &=&\frac{1}{4 v_{45}} \ (V^\dagger_{UD} U_2^* M_E^{\textrm{diag}} U^\dagger_2 - M_D^{\textrm{diag}}V^T_{UD})_{\alpha 1}\,,\\
(U^T Y^{\overline{5}} E)_{1 \beta} &=& \frac{1}{4 v_{45}} \ (U_2 M_E^{\textrm{diag}} - V^*_{UD} M_D^{\textrm{diag}}V^\dagger_{UD} U_2^*)_{1 \beta}\,.
\end{eqnarray}
The color triplet contribution to $p \rightarrow e_\delta^+ \pi^0$ accordingly reads 
\begin{eqnarray*}
\Gamma(p \rightarrow e_\delta^+ \pi^0) &=& \frac{m_p \alpha^2}{2^{18} \pi f_\pi^2v_{45}^4 m_\Delta^4}\left| (V^\dagger_{UD} U_2^* M_E^{diag} U_2^\dagger)_{11}-(V_{UD})_{11}m_d\right|^2\\&& \left| (U_2 M_E^{diag}-V^*_{UD} M_D^{diag} V^\dagger_{UD} U_2^*)_{1 \delta}\right|^2 (1+D+F)^2\,,
\end{eqnarray*}
where $v_{45}$ represents the VEV of the $45$-dimensional representation.
With the ansatz given in Eq.~\eqref{U2} we obtain the following expressions
\begin{eqnarray}
\Gamma(p \rightarrow e^+ \pi^0)&\approx&\frac{m_p \alpha^2}{2^{18} \pi f_\pi^2v_{45}^4 m_\Delta^4} 
\left|(V_{UD})_{13}\right|^2 m_b^2 m_\tau^2 (1+D+F)^2\,,\\
\Gamma(p \rightarrow \mu^+ \pi^0)&\approx&\frac{m_p \alpha^2}{2^{18} \pi f_\pi^2v_{45}^4 m_\Delta^4} 
\left|(V_{UD})_{12}\right|^2 m_s^2 m_\tau^2 (1+D+F)^2\,.
\end{eqnarray}
These should be compared with corresponding results for the exchange of the triplet in the  $5$-dimensional representation given in Eqs.~\eqref{A_1} and~\eqref{A_2} to obtain
\begin{eqnarray}
\label{ratio}
\left(\frac{\Gamma(p \rightarrow e_\delta^+ \pi^0)^5}{\Gamma(p \rightarrow e_\delta^+ \pi^0)^{45}}\right)_{max}=576\left(\frac{v_{45}}{v_5}\right)^4, \qquad \delta=1,2.
\end{eqnarray}
We see that the $5$-dimensional triplet dominates over the $45$-dimensional triplet for moderate values of $v_{45}$ if all other relevant parameters are the same.

\subsubsection{The neutral anti-lepton final state}
In order to incorporate $p \rightarrow \pi^+  \bar{\nu}$ and $p \rightarrow K^+ \bar{\nu}$ decay modes in our study we note that one is free to sum over the neutrino flavors in the final state. The relevant coefficients that enter widths for these decays are $a(d_{\alpha}, d_{\beta}, \nu_i)$ and $a (d_{\alpha}, d_{\beta}^C, \nu_i)$ when the exchanged state is the triplet in the  $5$-dimensional representation. To find them we need 
\begin{equation}
\label{X_1}
(D^T Y^{\overline{5}} N)_{\beta i} = -\frac{1}{2 v_5} \ (3 M_D^{\textrm{diag}} V^\dagger_{UD} U_2^* V_{EN}+V^T_{UD} U_2 M_E^{\textrm{diag}} V_{EN})_{\beta i}
\end{equation}
where the relevant sums yield
\begin{eqnarray}
\nonumber
&&\sum_i (D^T Y^{\overline{5}} N)_{\alpha i}(D^T Y^{\overline{5}} N)^*_{\beta i}=\\
&=&\frac{1}{4 v_5^2} \sum_j (3 M_D^{\textrm{diag}} V^\dagger_{UD} U_2^*+V^T_{UD}U_2 M_E^{\textrm{diag}})_{\alpha j} (3 M_D^{\textrm{diag}} V^T_{UD} U_2+V^\dagger_{UD}U^*_2 M_E^{\textrm{diag}})_{\beta j} \label{sum}
\end{eqnarray}
The upshot of these results is that widths for decays with neutral anti-lepton in the final state again depend only on $U_2$ as far as the mixing parameters are concerned. One can maximize amplitudes for $p \rightarrow K^+ \bar{\nu}$ and $p \rightarrow \pi^+ \bar{\nu}$ by taking the contributions proportional to $m_\tau$ in Eq.~\eqref{sum} and using the same ansatz for $U_2$ as before. We find that $\Gamma(p \rightarrow K^+ \bar{\nu})$ dominates over widths for proton decays into charged anti-leptons. It reads
\begin{equation}
\Gamma(p \rightarrow K^+ \bar{\nu}) \approx \frac{(m^2_p-m^2_{K^+})^2}{128 \pi f_\pi^2 m_p^3} \frac{\alpha^2 m_\tau^4}{v_5^4 m_\Delta^4}|V_{UD})_{12}|^2\left[1+\frac{m_p}{2 m_\Sigma}(D-F)+\frac{m_p}{2 m_\Lambda}(D+3F)\right]^2.
\end{equation}
This is an important result. It tells us that the most conservative bound on the scalar sector comes from the proton decays into neutral anti-leptons in the final state.  

If one compares the most conservative contributions of the triplets in the $5$- and $45$-dimensional representations towards $p \rightarrow K^+ \bar{\nu}$ one obtains
\begin{eqnarray}
\label{ratio_neutrino}
\left(\frac{\Gamma(p \rightarrow K^+ \bar{\nu})^5}{\Gamma(p \rightarrow K^+ \bar{\nu})^{45}}\right)_{max}=1024 \left(\frac{v_{45}}{v_5}\right)^4.
\end{eqnarray}
Again, the $5$-dimensional triplet dominates over the $45$-dimensional triplet for moderate values of $v_{45}$.

We can now numerically analyze all the decay modes given in Table~\ref{table:Proton_Decay_Bounds} to find the current bounds on the triplet mass in $SU(5)$ with symmetric Yukawa couplings. We take values of quark and lepton masses at $M_Z$, as given in~\cite{Dorsner:2006hw} , and neglect any running of the relevant
coefficients, for simplicity. These effects can be accounted for in a
straightforward manner. The CKM angles, when needed, are taken from Ref.~\cite{Nakamura:2010zzi}. We have randomly generated one million sets of values for nine parameters of $U_2$ unitary matrix and five phases of $V_{UD}$ to find the bounds presented in Table~\ref{table:Proton_Decay_MINMAX}. As it turns out, it is $p \rightarrow K^+ \bar{\nu}$ that dominates in all instances. With that in mind we can write that the most and least conservative bounds read
\begin{equation}
\label{DELTA_MASS_LIMIT}
m_{\Delta}>1.2 \times 10^{13} \left(\frac{\alpha}{0.0112\,\mathrm{GeV}^3}\right)^{1/2} \left(\frac{100\,\mathrm{GeV}}{v_5}\right)\,\mathrm{GeV}.
\end{equation}
\begin{equation}
\label{DELTA_MASS_LIMIT_a}
m_{\Delta}>1.5 \times 10^{11} \left(\frac{\alpha}{0.0112\,\mathrm{GeV}^3}\right)^{1/2} \left(\frac{100\,\mathrm{GeV}}{v_5}\right)\,\mathrm{GeV}.
\end{equation}

\begin{table}[h]
\begin{center}
\begin{tabular}{|l||r|r|}
\hline Channel & $m_\Delta$ (GeV) & $m_\Delta$ (GeV)\\
\hline 
$p \rightarrow \pi^0 e^+ $ & $1.9 \times 10^{10}$ & $4.9 \times 10^{12}$\\
$p \rightarrow \pi^0 \mu^+ $ & $2.8 \times 10^{10}$& $5.4 \times 10^{12}$\\
$p \rightarrow K^0 e^+$ & $1.7 \times 10^{10}$& $1.5 \times 10^{12}$\\
$p \rightarrow K^0 \mu^+$ & $2.0 \times 10^{10}$& $2.1 \times 10^{12}$\\
$p \rightarrow \eta e^+$ & $1.1\times 10^{10}$& $4.0\times 10^{11}$\\
$p \rightarrow \eta \mu^+$ & $7.2\times 10^{9}$& $2.7 \times 10^{11}$\\
$p \rightarrow \pi^+  \bar{\nu}$ & $2.2\times 10^{10}$& $8.0 \times 10^{12}$\\
$p \rightarrow K^+ \bar{\nu}$ & $1.5 \times 10^{11}$& $1.2 \times 10^{13}$\\
\hline
\end{tabular}
\end{center}
\caption{The least conservative  (second column) and the most conservative (third column) experimental lower bounds on triplet mass in the $5$-dimensional representation of $SU(5)$ with symmetric Yukawa couplings.}
\label{table:Proton_Decay_MINMAX}
\end{table}

To summarize, if one is to maximize contributions from the triplets in
the $5$- and $45$-dimensional representations towards proton decay
within renormalizable $SU(5)$ framework with symmetric mass matrices
the current bounds on the triplet mass scale are
given in Eqs.~\eqref{DELTA_MASS_LIMIT} and~\eqref{DELTA_MASS_LIMIT_a} if the color
triplet in $\bm{5}$ of Higgs dominates in the most and least conservative scenario, respectively. In other words, any 
$SU(5)$ scenario where the triplet scalar mass exceeds the most conservative bound of Eq.~\eqref{DELTA_MASS_LIMIT} is
certainly safe with regard to the proton decay constraints on the scalar mediated proton decay. If triplet
is to be lighter than that, one needs to explicitly check if
particular implementation of Yukawa couplings allows for such
scenario. The $5$-dimensional triplet dominance is determined through relations given in Eqs.~\eqref{ratio} and~\eqref{ratio_neutrino}. Finally, if the triplet mass is below the least conservative bound of Eq.~\eqref{DELTA_MASS_LIMIT_a} the $SU(5)$ model is not viable. 
\subsection{Color triplet in flipped $SU(5)$}

Flipped $SU(5)$ is well-known for the so-called missing partner mechanism that naturally addresses scalar mediated proton decay by making the triplet scalar in the $5$-dimensional representation heavy enough. Be that as it may, the operators associated with the triplet exchange can be suppressed with ease if $(U^T ( Y^{10}+Y^{10\,T}) D)_{1 \alpha}=0$ and $(D_C^{\dagger} Y^{\overline{5}\,*} U_C^*)_{\alpha 1}=0$, $\alpha=1,2$, in the Majorana neutrino case. Note that in flipped $SU(5)$ the Dirac mass matrix for neutrinos is proportional to the up-quark mass matrix. This implies that flipped $SU(5)$ predicts Majorana nature of neutrinos. 

The operator suppression can be implemented only if the Yukawa couplings are treated as free parameters. That is not the case in the minimal realistic version of flipped $SU(5)$ theory where Yukawa couplings are related to fermion masses and mixing parameters. In the minimal scenario it is sufficient to have only one $5$-dimensional scalar representation present to generate realistic charged fermion masses. We accordingly analyze predictions of a flipped $SU(5)$ scenario with a single color triplet state. To be able to compare the flipped $SU(5)$ results with the case of ordinary $SU(5)$ we again take $M_{U,D,E}=M_{U,D,E}^T$. 

\subsubsection{The charged anti-lepton final state}

We get the following result for the $p \rightarrow e_\delta^+ \pi^0$ ($\delta=1,2$) partial decay widths
\begin{equation}
\Gamma(p \rightarrow e_\delta^+ \pi^0) = \frac{m_p}{64 \pi f^2} \frac{\alpha^2}{v_5^4 m_\Delta^4}\left|(V_{UD})_{11}(m_d-m_u)\right|^2 4 (m_u^2+m_e^2)|(U_2)_{1\delta}|^2 (1+D+F)^2,
\end{equation}
where $v_5 (=\sqrt{2}\,246$\,GeV) represents the VEV of the $5$-dimensional representation. Interestingly enough, if $U_2$ takes the form given in Eq.~\eqref{U2} it would yield suppressed partial decay widths for $p \rightarrow e_\delta^+ \pi^0$, $\delta=1,2$. In other words, the setup that enhances partial proton decay rates with charged anti-lepton in the final state in $SU(5)$ framework suppresses corresponding rates in flipped $SU(5)$. 

If we want to be conservative with regard to the limit on $m_\Delta$ it is sufficient to maximize relevant decay widths. This can be done by taking $(U_2)_{1\delta}=1$ for $p \rightarrow e_\delta^+ \pi^0$ ($\delta=1,2$) to obtain
\begin{equation}
\label{DELTA_MASS_LIMIT_FSU5}
m_{\Delta}>3.6 \times 10^{10} \left(\frac{\alpha}{0.0112\,\mathrm{GeV}^3}\right)^{1/2}\,\mathrm{GeV}.
\end{equation}
This limit comes out to be significantly weaker with respect to the
corresponding limit we presented in the $SU(5)$ case.

\subsubsection{The neutral anti-lepton final state}
To find decay widths for $p \rightarrow \pi^+  \bar{\nu}$ and $p \rightarrow K^+ \bar{\nu}$ channels we need to determine the form of $a (d_{\alpha}, d_{\beta}, \nu_i)$ and $a (d_{\alpha}, d_{\beta}^C, \nu_i)$ coefficients. In the minimal model with symmetric mass matrices the relevant input reads
\begin{eqnarray}
(U^T(Y^{10}+Y^{10\,T}) D)_{1\alpha}&=& -\frac{1}{\sqrt{2} v_5} (V^*_{UD} M_D^{\textrm{diag}})_{1\alpha}\,,\\
(D^T Y^{\bar{5}} N)_{\beta i}&=& -\frac{2}{v_5} (V^T_{UD} M_U^{\textrm{diag}} U_2^* V_{EN})_{\beta i}\,,\\
(D^{\dagger} Y^{\overline{5}\,*} U^*)_{\beta 1}&=&-\frac{2}{v_5} (V^\dagger_{UD} M_U^{\textrm{diag}})_{\beta 1}\,,
\end{eqnarray}
with the following sum over neutrino flavors
\begin{equation}
\label{sum_b}
\sum_i (D^T Y^{\overline{5}} N)_{\alpha i}(D^T Y^{\overline{5}} N)^*_{\beta i}=\frac{4}{v_5^2} (V^T_{UD} (M_U^{\textrm{diag}})^2 V^T_{UD})_{\alpha \beta} . 
\end{equation}
The sum over neutrino flavors in the final state eliminates dependence on $U_2$---the matrix that represents mismatch between rotations in the quark and lepton sectors---leaving us with decay widths that depend only on known masses and mixing parameters. This makes minimal flipped $SU(5)$ with symmetric mass matrices rather unique. We find the following limit on the triplet mass that originates from experimental constraints on $p \rightarrow K^+ \bar{\nu}$ channel
\begin{equation}
\label{DELTA_MASS_LIMIT_b}
m_{\Delta}>1.0 \times 10^{12} \left(\frac{\alpha}{0.0112\,\mathrm{GeV}^3}\right)^{1/2}\,\mathrm{GeV}.
\end{equation}

The fact that $p \rightarrow \pi^+  \bar{\nu}$ is also a clean channel in the sense that it depends only on the CKM angles and known fermion masses means that the minimal flipped $SU(5)$ predicts ratio between $\Gamma(p \rightarrow \pi^+  \bar{\nu})$ and $\Gamma(p \rightarrow K^+  \bar{\nu})$. We find it to be
\begin{equation}
\label{ratio_neutrino_b}
\frac{\Gamma(p \rightarrow \pi^+ \bar{\nu})}{\Gamma(p \rightarrow K^+ \bar{\nu})}=9.0.
\end{equation}
This result is not sensitive to exact value of the nucleon matrix elements and running of relevant coefficient. It thus represents firm prediction within the framework of the minimal flipped $SU(5)$ with symmetric Yukawa couplings.

\section{Higher order contributions}
\label{Higher_Order}
In the $SU(5)$ framework the states $(\bar{\bm{3}},\bm{1},4/3)$ and
$\Delta^3 \in (\bm{3},\bm{3},-1/3)$ violate $B$ and $L$ and do not
contribute to dimension-six proton decay operators at tree-level.  Antisymmetry of
their Yukawa couplings to two up quarks only allows for dimension-six
operators involving $c$ or $t$ quarks that produce $B$ number
violation in charm or top decays~\cite{Dong:2011rh}, but these
operators do not affect the proton stability due to large masses of
$c$ and $t$ quarks.  However, an additional $W$ boson exchange opens
decay channels with final states that are kinematically accessible to
proton decay.

\subsection{Box mediated dimension-six operator
  from $(\bar{\bm{3}},\mathbf{1},4/3) \in 45$} 
\begin{figure}[!hbtp]
  \centering
  \begin{tabular}{cp{1cm}c}
      \includegraphics[width=0.33\textwidth]{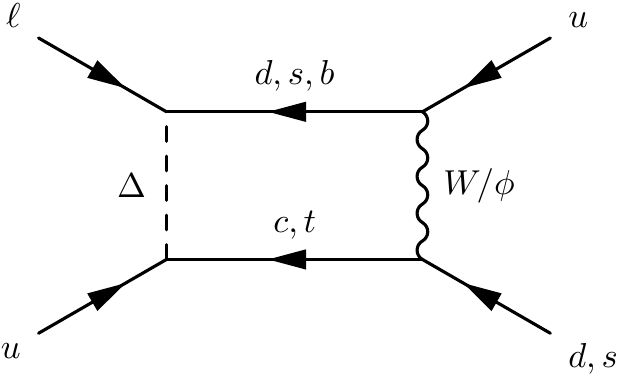} & & \includegraphics[width=0.33\textwidth]{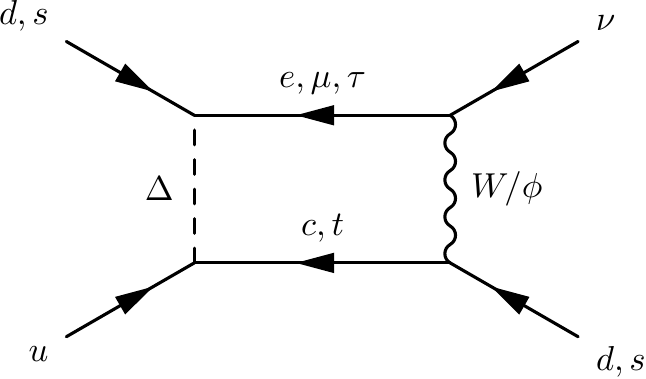}
  \end{tabular}
  \caption{Box diagrams with $(\bar{\bm{3}},\bm{1},4/3)$ state that generate
    $d=6$ operators of flavor $uud\ell$ and $udd\nu$.}
  \label{fig:boxes}
\end{figure}

One possibility is to make a box diagram with a single $W$ exchange
leading to the $d=6$ operator, as shown on Fig.~\ref{fig:boxes}.
In the literature, proton decay mediation involving $W$ boson
exchanges were considered in \cite{Baldes:2011mh,Hou:2005iu, Dong:2011rh}. We calculate the box diagram in the approximation
where we neglect external momenta, however, we keep both
virtual fermions massive since the right-handed $\Delta$ interactions
force chirality flips on internal fermion lines and thus the diagram
would vanish if both fermions were massless. Evaluation of the diagrams
with $W$ and would-be Goldstones leads to gauge invariant and finite
amplitude.  Then we find that $\Delta(\bar{\bm{3}}, \bm{1}, 4/3)$ generates
two effective coefficients:
\begin{eqnarray}
a(d_\alpha, e^C_\beta) 
   =-\frac{G_F}{4\pi^2 m_W^2} \sum_{j,k}&&\left[U_C^\dagger
     (Y^{10*}-Y^{10\dagger})U_C^*\right]_{1j} \left[D_C^\dagger
   Y^{\bar 5 \dagger} E_C^*\right]_{k\beta} \\
 &&\, m_{u_j} V_{j\alpha}\,
   m_{d_k} V_{uk}^* 
  \,J(x_\Delta,x_{u_j},x_{d_k})\,, \label{eq:D6uudl}\nonumber\\
a(d_\alpha, d^C_\beta,\nu_i)=-\frac{G_F}{4\pi^2 m_W^2} \sum_{j}&&\left[U_C^\dagger
     (Y^{10*}-Y^{10\dagger})U_C^*\right]_{1j} \left[D_C^\dagger
   Y^{\bar 5 \dagger} E_C^*\right]_{\beta i}\\
 &&\, m_{u_j} V_{j\alpha} \,m_{\ell_i} 
\, J(x_\Delta,x_{u_j},x_{\ell_i})\,.\label{eq:D6uddnu}\nonumber
\end{eqnarray}
Here, $V \equiv V_{CKM}$, while the leptonic mixing matrix
has been set to unity. (For the neutrino final states one would need to sum over all neutrino
  flavors.). Mass dependence, apart from helicity
flip factors, is encoded in function
$J$ (where $x_k \equiv m_k^2/m_W^2$)
\begin{eqnarray}
J(x,y,z) = &&\frac{(y -4) y  \log y}{(y -1)
     (y -x) (y-z)}+\frac{(z-4) z \log z}{(z-1) (z-y) (z-x
   )}\\
&&+\frac{(x-4) x \log x}{(x-1) (x -y)
   (x-z)}\,.\nonumber
\end{eqnarray}
There are two distinct regimes of dynamics in the box, depending on
the presence of $t$ quark in the loop. When $j=3$ we expand to leading
order in $x_{\ell_i}$, $x_{\ell_i} \ll 1$, and find
\begin{equation}
  \label{eq:Jtop}
  J(x_\Delta, x_t, x_{\ell_i}) = \frac{1}{x_\Delta - x_t}
  \left[\frac{x_\Delta-4}{x_\Delta-1}\log
    x_\Delta - \frac{x_t-4}{x_t-1}\log x_t \right]\,.
\end{equation}
When both fermions are light compared to $W$ the $J$ function takes the following form
\begin{equation}
  \label{eq:Jlight}
    J(x_\Delta, x_{u_j}, x_{\ell_i}) = \frac{1}{x_\Delta}
  \left[\frac{x_\Delta-4}{x_\Delta-1}\log
    x_\Delta
+\frac{4}{x_{u_j}-x_{\ell_i}}\left(x_{\ell_i} \log x_{\ell_i}-x_{u_j} \log x_{u_j}\right) \right]\,.
\end{equation}
Contributions of the up-quark Yukawa couplings are weighted
approximately by $m_{u_j} V_{jd}$ for $j=2,3$ that run in the
box. The large mass of the $t$ quark comes with small element
$V_{td}$ that makes this product of the same magnitude as $m_c V_{cd}$.
Similar cancellation between mass and CKM hierarchies occurs for the
down-quarks and the weights obey $m_d V_{ud}
\sim m_s V_{us} \sim m_b V_{ub}$.

The $(\bar{\bm{3}}, \bm{1}, 4/3)$ state has been identified as a
suitable candidate to explain the anomalous value of the muon magnetic
moment. One of the leptoquark couplings $(D_C^\dagger Y^{\bar 5
  \dagger} E_C^*)_{i2}$ between the muon and one of the down quarks
$d_i$ must be of the order $\sim 2$, while other two must be $\lesssim
10^{-3}$ to suppress contributions to other down-quark and charged
lepton observables~\cite{Dorsner:2011ai}. In addition, the CDF and
D$\O$ measurements of forward-backward asymmetry in $t\bar t$
production can be explained by a large diquark coupling $(U_C^\dagger
(Y^{10*}-Y^{10\dagger})U_C^*)_{31} \sim 2$ to $ut$ quark
pair~\cite{Dorsner:2009mq}. Additional couplings between $uc$ and $ct$
quark pairs are constrained by charm and top physics processes and
their upper bounds are of the order $10^{-1}$ and $10^{-2}$,
respectively~\cite{Dorsner:2010cu}. Both puzzles can be explained for
a mass of the leptoquark of around $400\,$GeV. However, for a light
mass and with the abovementioned two large couplings the proton would
decay much too quickly to a muon final state via dominant contribution
of $t$ quark and one of the down quarks in the box
(c.f. Eq.~\eqref{eq:D6uudl}). Therefore, one has to find a second
amplitude of equal magnitude and opposite phase in order to achieve
cancellation between the first and the second amplitude.  As explained
in the preceding paragraph, all down quarks in the box that couple to
external muon have similar weights that come from loop dynamics and
CKM factors. As a result, the hierarchy of $d_i$ contributions to the
amplitude follows very closely hierarchy of the leptoquark couplings
$(D_C^\dagger Y^{\bar 5 \dagger} E_C^*)_{i2}$ and the required
cancellation cannot take place between the different down-quarks in
the box. Likewise, cancellation between $c$ and $t$ quarks in the
box diagram cannot occur for similar reason. 

\subsection{Tree-level dimension-nine operator from $(\bar{\bm{3}},\mathbf{1},4/3) \in 45$}
\label{dim9}
\begin{figure}[htbp!]
 \centering
 \begin{tabular}{cp{1cm}c}
    \includegraphics[width=0.4\textwidth]{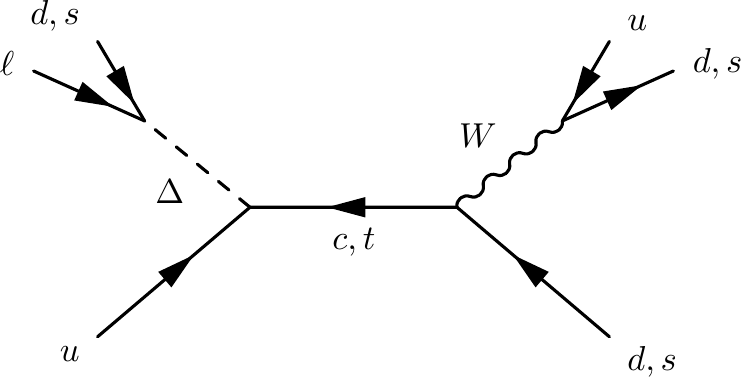}&&
 \includegraphics[width=0.4\textwidth]{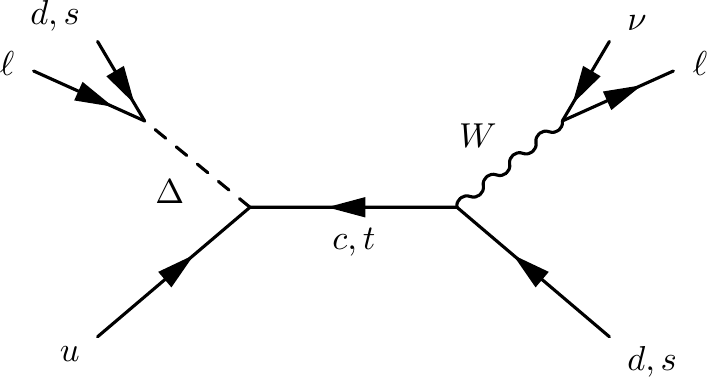}
 \end{tabular}
 \caption{$d=6$ proton decay operator induced by tree level
   $(\bar{\bm{3}},\bm{1},4/3) \in 45$ and $W$ exchanges.}
 \label{fig:dim9}
\end{figure}

The $W$ emission from the up-type quark leads to proton decay
amplitudes depicted on Fig.~\ref{fig:dim9}. Decays of this type have
been already mentioned in Ref.~\cite{Dong:2011rh}. We focus here on
the final state with a single charged lepton whose decay width is most
severely bounded experimentally. In this case the following $d=9$
effective operator is obtained
\begin{eqnarray}
 \label{eq:2}
 \mathcal{L}_9 = \sum_{U=c,t} &&\frac{-8 G_F V_{U\alpha} V^*_{u\gamma}} {m_{U} m_\Delta^2}  \left[U_C
    (Y^{10}-Y^{10T})^\dagger U_C^*\right]_{1U} \left[D_C^\dagger
  Y^{\bar 5 \dagger} E_C^*\right]_{\beta i}\nonumber\\
&&\,\epsilon_{abc}
(\overline{u_a^C} \gamma^\mu L d_{b\alpha})
(\overline{d_{c\beta}^C} R \ell_i ) 
(\overline{d_{k\gamma}} \gamma_\mu L u_k)
\,.
\end{eqnarray}
Here $U$ labels $c$ or $t$ quark, whereas external leptons $i,j = 1,2$
and down-type quarks $\alpha,\beta,\gamma = 1,2$, are all
light. $a,b,c,k$ are SU(3) color indices. $R(=(1+\gamma_5)/2)$ is the
right projection operator. We focus immediately on best constrained
channels, i.e., $p \to \pi^0 \ell_i^+$, and we set
$\alpha,\beta,\gamma = 1$.  The use of Fierz transformations leads to
the amplitude with scalar bilinears
\begin{eqnarray}
\label{eq:d9Fierzed}
{\cal M}_9^{p\to \pi^0 e_i^+}= \sum_{U=c,t}&& \frac{8 i G_F}{m_\Delta^2 m_{U}}  \left[U_C
    (Y^{10}-Y^{10T})^\dagger U_C^*\right]_{1U} \left[D_C^\dagger
  Y^{\bar 5 \dagger} E_C^*\right]_{1i}   V_{Ud} V_{ud}^* \\\,&&\epsilon_{abc} 
\Braket{\pi^0 \ell_i^+|  (\overline{\ell_i^C} R
   u_a)\,(\overline{d_k} R d_c)\,
 (\overline{u_k^C} L  d_b)  | p} + \textrm{tensor terms}\,.\nonumber
\label{e2bb}
\end{eqnarray}
One can estimate the above matrix element by employing the vacuum
saturation approximation. We insert the current $\overline{d_k} R
d_c$ between the vacuum and $\pi$ and end up with product of pion
creation and
proton annihilation amplitudes. The tensor terms which are invoked by
the Fierz relations in Eq.~\eqref{eq:d9Fierzed} cannot contribute in this case. The vacuum-to-pion
amplitude is
\begin{equation}
\Braket{\pi^0 | \overline{d_k} R d_c|0} = 
\frac{-i m_\pi^2  f_\pi}{4 \sqrt{2} m_d} \delta_{ck}\,,
\label{e2f}
\end{equation}
whereas the full amplitude is
\begin{eqnarray}
\label{e2g}
{\cal M}_9^{p\to \pi^0 e_i^+}= \sum_{U=c,t} &&\frac{-\sqrt{2}G_F}{m_\Delta^2 m_{U}}  \left[U_C
    (Y^{10}-Y^{10T})^\dagger U_C^*\right]_{1U} \left[D_C^\dagger
  Y^{\bar 5 \dagger} E_C^*\right]_{1i}   V_{Ud} V_{ud}^*
\frac{m_\pi^2 f_\pi}{m_d}\\ &&\,\epsilon_{abc}  
\Braket{\ell_i^+|   (\overline{u_a^C} L  d_b)\,(\overline{\ell_i^C} R
   u_c) | p}\nonumber\,.
\end{eqnarray}
The annihilation matrix element of the proton in~Eq.~(\ref{e2g}) has been most precisely evaluated
using lattice QCD~\cite{aoki}.  These authors have introduced
operators $O_{uds}^{\Gamma \Gamma'} = \epsilon_{abc} (\overline{u^C_a} 
\Gamma d_b ) \Gamma' s_c$ and defined constant $\alpha$ as
\begin{equation}
  \alpha R u_p = - \Braket{ 0| O_{udu}^{LR}|p}\,,
\label{e10}
\end{equation} 
where $u_p$ is the Dirac spinor of the proton. The recent value of
$\alpha$ obtained from lattice QCD calculation with domain wall
fermions~\cite{Aoki:2008ku} is $\alpha = -0.0112(25)\,\rm{GeV}^3$.  The decay
width is then
\begin{eqnarray}
\Gamma(p \to \pi^0 \ell_i^+) &= &\frac{G_F^2 f_\pi^2 m_\pi^4 \alpha^2}{16 \pi
 m_d^2} \frac{\lambda(m_p^2,m_{\ell_i}^2,m_\pi^2)^{1/2} (m_p^2+m_{\ell_i}^2-m_\pi^2)}{m_p^3}\\ &&\times\left|\sum_{U=c,t}
 \frac{V_{Ud} 
   \left[U_C
    (Y^{10}-Y^{10T})^\dagger U_C^*\right]_{1U} \left[D_C^\dagger
  Y^{\bar 5 \dagger} E_C^*\right]_{1i}}{m_U m_\Delta^2} \right|^2\,,\nonumber
\label{e2d}
\end{eqnarray}
where $\lambda(x,y,z) \equiv (x+y+z)^2-4(xy+yz+zx)$. From the
experimental limits in Tab.~\ref{table:Proton_Decay_Bounds} one obtains the following
bounds
\begin{eqnarray}
 \left|  \left[D_C^\dagger
  Y^{\bar 5 \dagger} E_C^*\right]_{11}\sum_{U=c,t}
 \frac{V_{U1}  \left[U_C
    (Y^{10}-Y^{10T})^\dagger U_C^*\right]_{1U}}{m_U} \right| &<& 2.4\times
10^{-20}\, \frac{m_\Delta^2}{(400\,\mathrm{GeV})^2} \,\mathrm{GeV}^{-1}\,,\\
 \left|  \left[D_C^\dagger
  Y^{\bar 5 \dagger} E_C^*\right]_{12} \sum_{U=c,t}
 \frac{V_{U1}  \left[U_C
    (Y^{10}-Y^{10T})^\dagger U_C^*\right]_{1U}}{m_U} \right| &<& 2.6\times
10^{-20}\, \frac{m_\Delta^2}{(400\,\mathrm{GeV})^2} \,\mathrm{GeV}^{-1}\,.
\end{eqnarray}
A comment is in order how phenomenologically preferred values of
leptoquark and diquark couplings cope with the above
constraints. Couplings to the electrons should be small and are in
particular not bounded from below, so the constraint from $\tau(p \to
\pi^0 e^+)$ can be avoided by putting $\left[D_C^\dagger Y^{\bar 5
    \dagger} E_C^*\right]_{11}$ effectively to zero. On the contrary,
low-energy leptoquark constraints, especially the $(g-2)_\mu$,
indicate that $\left[D_C^\dagger Y^{\bar 5 \dagger} E_C^*\right]_{12}$
could be large in some scenarios~\cite{Dorsner:2011ai}. In this case
we must require cancellation between the $c$ and $t$ quark amplitudes
that occurs when
\begin{equation}
\frac{\left[U_C (Y^{10}-Y^{10T})^\dagger U_C^*\right]_{12}}{ \left[U_C
    (Y^{10}-Y^{10T})^\dagger U_C^*\right]_{13}} \approx
-\frac{V_{td}}{V_{cd}} \frac{m_c}{m_t} \approx 2.7\times
10^{-4}\times e^{-0.37 i}\,.
\end{equation}
This can be achieved since the $\left[U_C (Y^{10}-Y^{10T})^\dagger
 U_C^*\right]_{12}$ is only bounded from above while at the same time
$\left[U_C (Y^{10}-Y^{10T})^\dagger U_C^*\right]_{13}$ is bounded from
below to satisfy observations in $t\bar t$
production. Finally, relative phase between the two couplings
can be freely adjusted since it is not probed by any experimental
observable to date.

Finally, for the state $(\bar{\bm{3}}, \bm{1}, -2/3)^{+6}$ present
in the flipped $SU(5)$ framework we can easily adapt the results
obtained above since the two states are indistinguishable at low
energies, provided we make the following substitutions
\begin{equation}
\left[U_C^\dagger (Y^{10*}-Y^{10\dagger})U_C^*\right] \to
\frac{1}{2} \left[U_C^\dagger Y^{\bar 5\dagger}U_C^*\right]\,,\qquad
\left[D_C^\dagger Y^{\bar 5 \dagger} E_C^*\right] \to - \left[D_C^\dagger  Y^{1 *} E_C^*\right]\,.
\end{equation}

To conclude, we note that despite the absence of the tree-level
contribution to proton decay of the $(\bar{\bm{3}},\bm{1},4/3)$ state,
weak corrections lead to proton destabilizing $d=6$ and $d=9$
operators. The effect of the $d=9$ operators can be rendered
adequately small even in the case of simultaneously large leptoquark
\emph{and} diquark couplings, a situation that is favored by
observables in $t\bar t$ production and value of $(g-2)_\mu$. This is
achieved by finely-tuned cancellation of two amplitudes. To the
contrary, similar cancellation is impossible in the case of $d=6$
operator for $p \to \pi^0 \mu^+$ decay and we are required to suppress
either all leptoquark couplings involving $\mu$ or all diquark
couplings. We conclude that the proton decay lifetime constraint
allows to fully address either $A_{FB}^{t\bar t}$ or $(g-2)_\mu$
observable with the $(\bar{\bm{3}},\bm{1},4/3)$ state, but not both.

\section{Conclusions}
\label{Conclusions}

Lepton and baryon number violating interactions are inherently present
within grand unified theories and are most severely constrained by the
observed proton stability. Proton decay can be mediated by vector or
scalar leptoquarks that violate both baryon and lepton number by one
unit. Vector leptoquarks that mediate proton decay have gauge
couplings to fermions and are not readily allowed to be far below the
unification scale. For the scalar leptoquarks, however, the freedom in
Yukawa couplings gives one more maneuverability to realize scenarios
with light scalar states. On the other hand, the very same Yukawas that are
responsible for proton decay very often need to account for the observed fermion mass
spectrum. An example of a setting with light leptoquark states was
presented in~\cite{Dorsner:2009cu,Dorsner:2010cu,Dorsner:2011ai} where
the low mass of the state $(\bar{\bm{3}}, \bm{1},4/3)$ had an impact
on low-energy flavor phenomenology. Most notably, it was found that by
tuning independently the two sets of Yukawa couplings, namely the
leptoquark and diquark Yukawas, one could reconcile the measured value
of forward-backward asymmetry in $t\bar t$ production and the value of
the magnetic moment of muon.

In this work, we have classified the scalar leptoquarks present in
$SU(5)$ and flipped $SU(5)$ grand unification frameworks that
mediate proton decay. In both frameworks the considered leptoquark
states reside in scalar representations of $SU(5)$ of dimension $5$,
$10$, $45$, or $50$. We integrate out the above states at tree-level
and parameterize their contributions in terms of effective
coefficients of a complete set of dimension-six effective operators.
The mass constraint on the color triplet state contained in the
5- and 45-dimensional representations is then derived. The precise lower
bound depends on the value of the vacuum expectation values of these
representation. For the vacuum expectation value of $100\,$GeV 
the least (most) conservative lower bound on the triplet mass that originates from the $p \rightarrow K^+ \bar{\nu}$ channel is approximately $10^{11}$\,GeV ($10^{13}$\,GeV). The corresponding bound is derived within the flipped $SU(5)$ framework to read $10^{12}$\,GeV and proves to be mixing independent. Moreover, the minimal flipped $SU(5)$ theory with symmetric mass matrices predicts $\Gamma(p \rightarrow \pi^+ \bar{\nu})/\Gamma(p \rightarrow K^+ \bar{\nu})=9$.

The two leptoquark states that do not contribute to proton decay at
tree-level are $(\bar{\bm{3}},\bm{1},4/3)$ and $(\bar{\bm{3}}, \bm{1},
-2/3)^{+6}$ in the standard and flipped $SU(5)$ frameworks,
respectively. We have estimated their contribution to dimension-six
operators via box diagram and the tree-level contribution to
dimension-nine operators. For the $(\bar{\bm{3}},\bm{1},4/3)$ state it
has been found that if it is to explain both the anomalous magnetic moment
of the muon and the $t \bar t$ forward-backward asymmetry, then the
contribution of the dimension-six operator would destabilize the proton
in $p \to \mu^+ \pi^0$ channel. Therefore only one of the two puzzles
can be addressed with this leptoquark state.

Light scalar leptoquarks can be either produced in pairs or in
association with SM fermions at the LHC and are a subject of
leptoquark and diquark resonance
searches~\cite{Giudice:2011ak,Aad:2012cy,Exotica,Chatrchyan:2011ns}.
To conclude, we can expect to find signals of these leptoquark states
at the LHC, although it seems very unlikely, in light of the
constraints from the proton lifetime measurements, that they would be
observed in a baryon number violating processes.
\begin{acknowledgments}
  We acknowledge enlightening discussions with Damir Be\v{c}irevi\'c
  and Jernej F. Kamenik. This work is supported in part by the
  Slovenian Research Agency. I.D. thanks the members of the ``Jo\v zef
  Stefan" Institute, where part of this work was completed, for their
  hospitality.
\end{acknowledgments}

\bibliography{refs}

\begin{thebibliography}{53}
\expandafter\ifx\csname natexlab\endcsname\relax\def\natexlab#1{#1}\fi
\expandafter\ifx\csname bibnamefont\endcsname\relax
  \def\bibnamefont#1{#1}\fi
\expandafter\ifx\csname bibfnamefont\endcsname\relax
  \def\bibfnamefont#1{#1}\fi
\expandafter\ifx\csname citenamefont\endcsname\relax
  \def\citenamefont#1{#1}\fi
\expandafter\ifx\csname url\endcsname\relax
  \def\url#1{\texttt{#1}}\fi
\expandafter\ifx\csname urlprefix\endcsname\relax\def\urlprefix{URL }\fi
\providecommand{\bibinfo}[2]{#2}
\providecommand{\eprint}[2][]{\url{#2}}


\bibitem[{\citenamefont{Aad et~al.}(2011)}]{Aad:2011uv}
\bibinfo{author}{\bibfnamefont{G.}~\bibnamefont{Aad}} \bibnamefont{et~al.}
  (\bibinfo{collaboration}{ATLAS Collaboration}), \bibinfo{journal}{Phys.Rev.}
  \textbf{\bibinfo{volume}{D83}}, \bibinfo{pages}{112006}
  (\bibinfo{year}{2011}), \eprint{1104.4481}.

\bibitem[{\citenamefont{Chatrchyan
  et~al.}(2011{\natexlab{a}})}]{Chatrchyan:2011ar}
\bibinfo{author}{\bibfnamefont{S.}~\bibnamefont{Chatrchyan}}
  \bibnamefont{et~al.} (\bibinfo{collaboration}{CMS Collaboration}),
  \bibinfo{journal}{Phys.Lett.} \textbf{\bibinfo{volume}{B703}},
  \bibinfo{pages}{246} (\bibinfo{year}{2011}{\natexlab{a}}),
  \eprint{1105.5237}.

\bibitem[{\citenamefont{Abazov et~al.}(2011{\natexlab{a}})}]{Abazov:2011qj}
\bibinfo{author}{\bibfnamefont{V.~M.} \bibnamefont{Abazov}}
  \bibnamefont{et~al.} (\bibinfo{collaboration}{D0 Collaboration}),
  \bibinfo{journal}{Phys.Rev.} \textbf{\bibinfo{volume}{D84}},
  \bibinfo{pages}{071104} (\bibinfo{year}{2011}{\natexlab{a}}),
  \eprint{1107.1849}.

\bibitem[{\citenamefont{Aaron et~al.}(2011)\citenamefont{Aaron, Alexa, Andreev,
  Backovic, Baghdasaryan et~al.}}]{Collaboration:2011qaa}
\bibinfo{author}{\bibfnamefont{F.}~\bibnamefont{Aaron}},
  \bibinfo{author}{\bibfnamefont{C.}~\bibnamefont{Alexa}},
  \bibinfo{author}{\bibfnamefont{V.}~\bibnamefont{Andreev}},
  \bibinfo{author}{\bibfnamefont{S.}~\bibnamefont{Backovic}},
  \bibinfo{author}{\bibfnamefont{A.}~\bibnamefont{Baghdasaryan}},
  \bibnamefont{et~al.}, \bibinfo{journal}{Phys.Lett.}
  \textbf{\bibinfo{volume}{B704}}, \bibinfo{pages}{388} (\bibinfo{year}{2011}),
  \eprint{1107.3716}.

\bibitem[{\citenamefont{Aad et~al.}(2012{\natexlab{a}})}]{Aad:2011ch}
\bibinfo{author}{\bibfnamefont{G.}~\bibnamefont{Aad}} \bibnamefont{et~al.}
  (\bibinfo{collaboration}{ATLAS Collaboration}), \bibinfo{journal}{Phys.Lett.}
  \textbf{\bibinfo{volume}{B709}}, \bibinfo{pages}{158}
  (\bibinfo{year}{2012}{\natexlab{a}}), \eprint{1112.4828}.

\bibitem[{\citenamefont{Aad et~al.}(2012{\natexlab{b}})}]{Aad:2012cy}
\bibinfo{author}{\bibfnamefont{G.}~\bibnamefont{Aad}} \bibnamefont{et~al.}
  (\bibinfo{collaboration}{ATLAS Collaboration})
  (\bibinfo{year}{2012}{\natexlab{b}}), \eprint{1203.3172}.

\bibitem[{\citenamefont{Leurer}(1994{\natexlab{a}})}]{Leurer:1993em}
\bibinfo{author}{\bibfnamefont{M.}~\bibnamefont{Leurer}},
  \bibinfo{journal}{Phys.Rev.} \textbf{\bibinfo{volume}{D49}},
  \bibinfo{pages}{333} (\bibinfo{year}{1994}{\natexlab{a}}),
  \eprint{hep-ph/9309266}.

\bibitem[{\citenamefont{Leurer}(1994{\natexlab{b}})}]{Leurer:1993qx}
\bibinfo{author}{\bibfnamefont{M.}~\bibnamefont{Leurer}},
  \bibinfo{journal}{Phys.Rev.} \textbf{\bibinfo{volume}{D50}},
  \bibinfo{pages}{536} (\bibinfo{year}{1994}{\natexlab{b}}),
  \eprint{hep-ph/9312341}.

\bibitem[{\citenamefont{Carpentier and Davidson}(2010)}]{Carpentier:2010ue}
\bibinfo{author}{\bibfnamefont{M.}~\bibnamefont{Carpentier}} \bibnamefont{and}
  \bibinfo{author}{\bibfnamefont{S.}~\bibnamefont{Davidson}},
  \bibinfo{journal}{Eur.Phys.J.} \textbf{\bibinfo{volume}{C70}},
  \bibinfo{pages}{1071} (\bibinfo{year}{2010}), \eprint{1008.0280}.

\bibitem[{\citenamefont{Saha et~al.}(2010)\citenamefont{Saha, Misra, and
  Kundu}}]{Saha:2010vw}
\bibinfo{author}{\bibfnamefont{J.~P.} \bibnamefont{Saha}},
  \bibinfo{author}{\bibfnamefont{B.}~\bibnamefont{Misra}}, \bibnamefont{and}
  \bibinfo{author}{\bibfnamefont{A.}~\bibnamefont{Kundu}},
  \bibinfo{journal}{Phys.Rev.} \textbf{\bibinfo{volume}{D81}},
  \bibinfo{pages}{095011} (\bibinfo{year}{2010}), \eprint{1003.1384}.

\bibitem[{\citenamefont{Dighe et~al.}(2010)\citenamefont{Dighe, Kundu, and
  Nandi}}]{Dighe:2010nj}
\bibinfo{author}{\bibfnamefont{A.}~\bibnamefont{Dighe}},
  \bibinfo{author}{\bibfnamefont{A.}~\bibnamefont{Kundu}}, \bibnamefont{and}
  \bibinfo{author}{\bibfnamefont{S.}~\bibnamefont{Nandi}},
  \bibinfo{journal}{Phys.Rev.} \textbf{\bibinfo{volume}{D82}},
  \bibinfo{pages}{031502} (\bibinfo{year}{2010}), \eprint{1005.4051}.

\bibitem[{\citenamefont{Weinberg}(1979)}]{weinberg1}
\bibinfo{author}{\bibfnamefont{S.}~\bibnamefont{Weinberg}},
  \bibinfo{journal}{Phys. Rev. Lett.} \textbf{\bibinfo{volume}{43}},
  \bibinfo{pages}{1566} (\bibinfo{year}{1979}).

\bibitem[{\citenamefont{Wilczek and Zee}(1979)}]{wilczek}
\bibinfo{author}{\bibfnamefont{F.}~\bibnamefont{Wilczek}} \bibnamefont{and}
  \bibinfo{author}{\bibfnamefont{A.}~\bibnamefont{Zee}},
  \bibinfo{journal}{Phys.Rev.Lett.} \textbf{\bibinfo{volume}{43}},
  \bibinfo{pages}{1571} (\bibinfo{year}{1979}).

\bibitem[{\citenamefont{Weinberg}(1980)}]{weinberg2}
\bibinfo{author}{\bibfnamefont{S.}~\bibnamefont{Weinberg}},
  \bibinfo{journal}{Phys. Rev.} \textbf{\bibinfo{volume}{D22}},
  \bibinfo{pages}{1694} (\bibinfo{year}{1980}).

\bibitem[{\citenamefont{Dorsner et~al.}(2009)\citenamefont{Dorsner, Fajfer,
  Kamenik, and Kosnik}}]{Dorsner:2009cu}
\bibinfo{author}{\bibfnamefont{I.}~\bibnamefont{Dorsner}},
  \bibinfo{author}{\bibfnamefont{S.}~\bibnamefont{Fajfer}},
  \bibinfo{author}{\bibfnamefont{J.~F.} \bibnamefont{Kamenik}},
  \bibnamefont{and} \bibinfo{author}{\bibfnamefont{N.}~\bibnamefont{Kosnik}},
  \bibinfo{journal}{Phys.Lett.} \textbf{\bibinfo{volume}{B682}},
  \bibinfo{pages}{67} (\bibinfo{year}{2009}), \eprint{0906.5585}.

\bibitem[{\citenamefont{Georgi and Glashow}(1974)}]{Georgi:1974sy}
\bibinfo{author}{\bibfnamefont{H.}~\bibnamefont{Georgi}} \bibnamefont{and}
  \bibinfo{author}{\bibfnamefont{S.}~\bibnamefont{Glashow}},
  \bibinfo{journal}{Phys.Rev.Lett.} \textbf{\bibinfo{volume}{32}},
  \bibinfo{pages}{438} (\bibinfo{year}{1974}).

\bibitem[{\citenamefont{De~Rujula et~al.}(1980)\citenamefont{De~Rujula, Georgi,
  and Glashow}}]{DeRujula:1980qc}
\bibinfo{author}{\bibfnamefont{A.}~\bibnamefont{De~Rujula}},
  \bibinfo{author}{\bibfnamefont{H.}~\bibnamefont{Georgi}}, \bibnamefont{and}
  \bibinfo{author}{\bibfnamefont{S.}~\bibnamefont{Glashow}},
  \bibinfo{journal}{Phys.Rev.Lett.} \textbf{\bibinfo{volume}{45}},
  \bibinfo{pages}{413} (\bibinfo{year}{1980}).

\bibitem[{\citenamefont{Georgi et~al.}(1981)\citenamefont{Georgi, Glashow, and
  Machacek}}]{Georgi:1980pw}
\bibinfo{author}{\bibfnamefont{H.}~\bibnamefont{Georgi}},
  \bibinfo{author}{\bibfnamefont{S.}~\bibnamefont{Glashow}}, \bibnamefont{and}
  \bibinfo{author}{\bibfnamefont{M.}~\bibnamefont{Machacek}},
  \bibinfo{journal}{Phys.Rev.} \textbf{\bibinfo{volume}{D23}},
  \bibinfo{pages}{783} (\bibinfo{year}{1981}).

\bibitem[{\citenamefont{Barr}(1982)}]{Barr:1981qv}
\bibinfo{author}{\bibfnamefont{S.~M.} \bibnamefont{Barr}},
  \bibinfo{journal}{Phys.Lett.} \textbf{\bibinfo{volume}{B112}},
  \bibinfo{pages}{219} (\bibinfo{year}{1982}).

\bibitem[{\citenamefont{Fileviez~Perez}(2004)}]{FileviezPerez:2004hn}
\bibinfo{author}{\bibfnamefont{P.}~\bibnamefont{Fileviez~Perez}},
  \bibinfo{journal}{Phys.Lett.} \textbf{\bibinfo{volume}{B595}},
  \bibinfo{pages}{476} (\bibinfo{year}{2004}), \eprint{hep-ph/0403286}.

\bibitem[{\citenamefont{Dorsner and
  Fileviez~Perez}(2005{\natexlab{a}})}]{Dorsner:2004xx}
\bibinfo{author}{\bibfnamefont{I.}~\bibnamefont{Dorsner}} \bibnamefont{and}
  \bibinfo{author}{\bibfnamefont{P.}~\bibnamefont{Fileviez~Perez}},
  \bibinfo{journal}{Phys.Lett.} \textbf{\bibinfo{volume}{B605}},
  \bibinfo{pages}{391} (\bibinfo{year}{2005}{\natexlab{a}}),
  \eprint{hep-ph/0409095}.

\bibitem[{\citenamefont{Dorsner and
  Fileviez~Perez}(2005{\natexlab{b}})}]{Dorsner:2004jj}
\bibinfo{author}{\bibfnamefont{I.}~\bibnamefont{Dorsner}} \bibnamefont{and}
  \bibinfo{author}{\bibfnamefont{P.}~\bibnamefont{Fileviez~Perez}},
  \bibinfo{journal}{Phys.Lett.} \textbf{\bibinfo{volume}{B606}},
  \bibinfo{pages}{367} (\bibinfo{year}{2005}{\natexlab{b}}),
  \eprint{hep-ph/0409190}.

\bibitem[{\citenamefont{Dorsner and
  Fileviez~Perez}(2005{\natexlab{c}})}]{Dorsner:2004xa}
\bibinfo{author}{\bibfnamefont{I.}~\bibnamefont{Dorsner}} \bibnamefont{and}
  \bibinfo{author}{\bibfnamefont{P.}~\bibnamefont{Fileviez~Perez}},
  \bibinfo{journal}{Phys.Lett.} \textbf{\bibinfo{volume}{B625}},
  \bibinfo{pages}{88} (\bibinfo{year}{2005}{\natexlab{c}}),
  \eprint{hep-ph/0410198}.

\bibitem[{\citenamefont{Nath and Fileviez~Perez}(2007)}]{Nath:2006ut}
\bibinfo{author}{\bibfnamefont{P.}~\bibnamefont{Nath}} \bibnamefont{and}
  \bibinfo{author}{\bibfnamefont{P.}~\bibnamefont{Fileviez~Perez}},
  \bibinfo{journal}{Phys.Rept.} \textbf{\bibinfo{volume}{441}},
  \bibinfo{pages}{191} (\bibinfo{year}{2007}), \eprint{hep-ph/0601023}.

\bibitem[{\citenamefont{Fileviez~Perez}(2007)}]{Perez:2007rm}
\bibinfo{author}{\bibfnamefont{P.}~\bibnamefont{Fileviez~Perez}},
  \bibinfo{journal}{Phys. Lett.} \textbf{\bibinfo{volume}{B654}},
  \bibinfo{pages}{189} (\bibinfo{year}{2007}), \eprint{hep-ph/0702287}.

\bibitem[{\citenamefont{CDF}(2011)}]{cdfttbar}
\bibinfo{author}{\bibnamefont{CDF}}, \bibinfo{journal}{Public note 10584}
  (\bibinfo{year}{2011}).

\bibitem[{\citenamefont{Abazov et~al.}(2011{\natexlab{b}})}]{Abazov:2011rq}
\bibinfo{author}{\bibfnamefont{V.~M.} \bibnamefont{Abazov}}
  \bibnamefont{et~al.} (\bibinfo{collaboration}{D0 Collaboration}),
  \bibinfo{journal}{Phys.Rev.} \textbf{\bibinfo{volume}{D84}},
  \bibinfo{pages}{112005} (\bibinfo{year}{2011}{\natexlab{b}}),
  \eprint{1107.4995}.

\bibitem[{\citenamefont{Dorsner
  et~al.}(2010{\natexlab{a}})\citenamefont{Dorsner, Fajfer, Kamenik, and
  Kosnik}}]{Dorsner:2009mq}
\bibinfo{author}{\bibfnamefont{I.}~\bibnamefont{Dorsner}},
  \bibinfo{author}{\bibfnamefont{S.}~\bibnamefont{Fajfer}},
  \bibinfo{author}{\bibfnamefont{J.~F.} \bibnamefont{Kamenik}},
  \bibnamefont{and} \bibinfo{author}{\bibfnamefont{N.}~\bibnamefont{Kosnik}},
  \bibinfo{journal}{Phys.Rev.} \textbf{\bibinfo{volume}{D81}},
  \bibinfo{pages}{055009} (\bibinfo{year}{2010}{\natexlab{a}}),
  \eprint{0912.0972}.

\bibitem[{\citenamefont{Del~Nobile et~al.}(2010)\citenamefont{Del~Nobile,
  Franceschini, Pappadopulo, and Strumia}}]{DelNobile:2009st}
\bibinfo{author}{\bibfnamefont{E.}~\bibnamefont{Del~Nobile}},
  \bibinfo{author}{\bibfnamefont{R.}~\bibnamefont{Franceschini}},
  \bibinfo{author}{\bibfnamefont{D.}~\bibnamefont{Pappadopulo}},
  \bibnamefont{and} \bibinfo{author}{\bibfnamefont{A.}~\bibnamefont{Strumia}},
  \bibinfo{journal}{Nucl.Phys.} \textbf{\bibinfo{volume}{B826}},
  \bibinfo{pages}{217} (\bibinfo{year}{2010}), \eprint{0908.1567}.

\bibitem[{\citenamefont{Vecchi}(2011)}]{Vecchi:2011ab}
\bibinfo{author}{\bibfnamefont{L.}~\bibnamefont{Vecchi}},
  \bibinfo{journal}{JHEP} \textbf{\bibinfo{volume}{1110}}, \bibinfo{pages}{003}
  (\bibinfo{year}{2011}), \bibinfo{note}{references added, published version},
  \eprint{1107.2933}.

\bibitem[{\citenamefont{Dorsner et~al.}(2011)\citenamefont{Dorsner, Drobnak,
  Fajfer, Kamenik, and Kosnik}}]{Dorsner:2011ai}
\bibinfo{author}{\bibfnamefont{I.}~\bibnamefont{Dorsner}},
  \bibinfo{author}{\bibfnamefont{J.}~\bibnamefont{Drobnak}},
  \bibinfo{author}{\bibfnamefont{S.}~\bibnamefont{Fajfer}},
  \bibinfo{author}{\bibfnamefont{J.~F.} \bibnamefont{Kamenik}},
  \bibnamefont{and} \bibinfo{author}{\bibfnamefont{N.}~\bibnamefont{Kosnik}},
  \bibinfo{journal}{JHEP} \textbf{\bibinfo{volume}{1111}}, \bibinfo{pages}{002}
  (\bibinfo{year}{2011}), \eprint{1107.5393}.

\bibitem[{\citenamefont{Dorsner
  et~al.}(2010{\natexlab{b}})\citenamefont{Dorsner, Fajfer, Kamenik, and
  Kosnik}}]{Dorsner:2010cu}
\bibinfo{author}{\bibfnamefont{I.}~\bibnamefont{Dorsner}},
  \bibinfo{author}{\bibfnamefont{S.}~\bibnamefont{Fajfer}},
  \bibinfo{author}{\bibfnamefont{J.~F.} \bibnamefont{Kamenik}},
  \bibnamefont{and} \bibinfo{author}{\bibfnamefont{N.}~\bibnamefont{Kosnik}},
  \bibinfo{journal}{Phys.Rev.} \textbf{\bibinfo{volume}{D82}},
  \bibinfo{pages}{094015} (\bibinfo{year}{2010}{\natexlab{b}}),
  \eprint{1007.2604}.

\bibitem[{\citenamefont{Bajc and Senjanovic}(2007)}]{Bajc:2006ia}
\bibinfo{author}{\bibfnamefont{B.}~\bibnamefont{Bajc}} \bibnamefont{and}
  \bibinfo{author}{\bibfnamefont{G.}~\bibnamefont{Senjanovic}},
  \bibinfo{journal}{JHEP} \textbf{\bibinfo{volume}{0708}}, \bibinfo{pages}{014}
  (\bibinfo{year}{2007}), \eprint{hep-ph/0612029}.

\bibitem[{\citenamefont{Oshimo}(2009)}]{Oshimo:2009ia}
\bibinfo{author}{\bibfnamefont{N.}~\bibnamefont{Oshimo}},
  \bibinfo{journal}{Phys.Rev.} \textbf{\bibinfo{volume}{D80}},
  \bibinfo{pages}{075011} (\bibinfo{year}{2009}), \eprint{0907.3400}.

\bibitem[{\citenamefont{Slansky}(1981)}]{Slansky:1981yr}
\bibinfo{author}{\bibfnamefont{R.}~\bibnamefont{Slansky}},
  \bibinfo{journal}{Phys.Rept.} \textbf{\bibinfo{volume}{79}},
  \bibinfo{pages}{1} (\bibinfo{year}{1981}).

\bibitem[{\citenamefont{Georgi and Jarlskog}(1979)}]{Georgi:1979df}
\bibinfo{author}{\bibfnamefont{H.}~\bibnamefont{Georgi}} \bibnamefont{and}
  \bibinfo{author}{\bibfnamefont{C.}~\bibnamefont{Jarlskog}},
  \bibinfo{journal}{Phys.Lett.} \textbf{\bibinfo{volume}{B86}},
  \bibinfo{pages}{297} (\bibinfo{year}{1979}).

\bibitem[{\citenamefont{Miura}(2010)}]{Miura:2010zz}
\bibinfo{author}{\bibfnamefont{M.}~\bibnamefont{Miura}}, \bibinfo{journal}{PoS}
  \textbf{\bibinfo{volume}{ICHEP2010}}, \bibinfo{pages}{408}
  (\bibinfo{year}{2010}).

\bibitem[{\citenamefont{Miura}(2011)}]{SuperKNew}
\bibinfo{author}{\bibfnamefont{M.}~\bibnamefont{Miura}}
  (\bibinfo{publisher}{Presented at the 2011 Workshop on Baryon \& Lepton
  Number Violation, Gatlinburg, Tennesee}, \bibinfo{year}{2011}).

\bibitem[{\citenamefont{Kobayashi et~al.}(2005)}]{Kobayashi:2005pe}
\bibinfo{author}{\bibfnamefont{K.}~\bibnamefont{Kobayashi}}
  \bibnamefont{et~al.} (\bibinfo{collaboration}{Super-Kamiokande
  Collaboration}), \bibinfo{journal}{Phys.Rev.} \textbf{\bibinfo{volume}{D72}},
  \bibinfo{pages}{052007} (\bibinfo{year}{2005}), \eprint{hep-ex/0502026}.

\bibitem[{\citenamefont{Nishino et~al.}(2012)\citenamefont{Nishino, Abe,
  Hayato, Iida, Ikeda et~al.}}]{Nishino:2012rv}
\bibinfo{author}{\bibfnamefont{H.}~\bibnamefont{Nishino}},
  \bibinfo{author}{\bibfnamefont{K.}~\bibnamefont{Abe}},
  \bibinfo{author}{\bibfnamefont{Y.}~\bibnamefont{Hayato}},
  \bibinfo{author}{\bibfnamefont{T.}~\bibnamefont{Iida}},
  \bibinfo{author}{\bibfnamefont{M.}~\bibnamefont{Ikeda}}, \bibnamefont{et~al.}
  (\bibinfo{year}{2012}), \eprint{1203.4030}.

\bibitem[{\citenamefont{Nakamura et~al.}(2010)}]{Nakamura:2010zzi}
\bibinfo{author}{\bibfnamefont{K.}~\bibnamefont{Nakamura}} \bibnamefont{et~al.}
  (\bibinfo{collaboration}{Particle Data Group}), \bibinfo{journal}{J.Phys.G}
  \textbf{\bibinfo{volume}{G37}}, \bibinfo{pages}{075021}
  (\bibinfo{year}{2010}).

\bibitem[{\citenamefont{Claudson et~al.}(1982)\citenamefont{Claudson, Wise, and
  Hall}}]{Claudson:1981gh}
\bibinfo{author}{\bibfnamefont{M.}~\bibnamefont{Claudson}},
  \bibinfo{author}{\bibfnamefont{M.~B.} \bibnamefont{Wise}}, \bibnamefont{and}
  \bibinfo{author}{\bibfnamefont{L.~J.} \bibnamefont{Hall}},
  \bibinfo{journal}{Nucl.Phys.} \textbf{\bibinfo{volume}{B195}},
  \bibinfo{pages}{297} (\bibinfo{year}{1982}).

\bibitem[{\citenamefont{Aoki et~al.}(2008)}]{Aoki:2008ku}
\bibinfo{author}{\bibfnamefont{Y.}~\bibnamefont{Aoki}} \bibnamefont{et~al.}
  (\bibinfo{collaboration}{RBC-UKQCD Collaboration}),
  \bibinfo{journal}{Phys.Rev.} \textbf{\bibinfo{volume}{D78}},
  \bibinfo{pages}{054505} (\bibinfo{year}{2008}), \eprint{0806.1031}.

\bibitem[{\citenamefont{Dorsner et~al.}(2007)\citenamefont{Dorsner,
  Fileviez~Perez, and Rodrigo}}]{Dorsner:2006hw}
\bibinfo{author}{\bibfnamefont{I.}~\bibnamefont{Dorsner}},
  \bibinfo{author}{\bibfnamefont{P.}~\bibnamefont{Fileviez~Perez}},
  \bibnamefont{and} \bibinfo{author}{\bibfnamefont{G.}~\bibnamefont{Rodrigo}},
  \bibinfo{journal}{Phys.Rev.} \textbf{\bibinfo{volume}{D75}},
  \bibinfo{pages}{125007} (\bibinfo{year}{2007}), \eprint{hep-ph/0607208}.

\bibitem[{\citenamefont{Dong et~al.}(2012)\citenamefont{Dong, Durieux, Gerard,
  Han, and Maltoni}}]{Dong:2011rh}
\bibinfo{author}{\bibfnamefont{Z.}~\bibnamefont{Dong}},
  \bibinfo{author}{\bibfnamefont{G.}~\bibnamefont{Durieux}},
  \bibinfo{author}{\bibfnamefont{J.-M.} \bibnamefont{Gerard}},
  \bibinfo{author}{\bibfnamefont{T.}~\bibnamefont{Han}}, \bibnamefont{and}
  \bibinfo{author}{\bibfnamefont{F.}~\bibnamefont{Maltoni}},
  \bibinfo{journal}{Phys.Rev.} \textbf{\bibinfo{volume}{D85}},
  \bibinfo{pages}{016006} (\bibinfo{year}{2012}), \bibinfo{note}{5 pages, 3
  figures}, \eprint{1107.3805}.

\bibitem[{\citenamefont{Baldes et~al.}(2011)\citenamefont{Baldes, Bell, and
  Volkas}}]{Baldes:2011mh}
\bibinfo{author}{\bibfnamefont{I.}~\bibnamefont{Baldes}},
  \bibinfo{author}{\bibfnamefont{N.~F.} \bibnamefont{Bell}}, \bibnamefont{and}
  \bibinfo{author}{\bibfnamefont{R.~R.} \bibnamefont{Volkas}},
  \bibinfo{journal}{Phys.Rev.} \textbf{\bibinfo{volume}{D84}},
  \bibinfo{pages}{115019} (\bibinfo{year}{2011}), \bibinfo{note}{10 pages, 9
  figures, references added, some typos corrected}, \eprint{1110.4450}.

\bibitem[{\citenamefont{Hou et~al.}(2005)\citenamefont{Hou, Nagashima, and
  Soddu}}]{Hou:2005iu}
\bibinfo{author}{\bibfnamefont{W.-S.} \bibnamefont{Hou}},
  \bibinfo{author}{\bibfnamefont{M.}~\bibnamefont{Nagashima}},
  \bibnamefont{and} \bibinfo{author}{\bibfnamefont{A.}~\bibnamefont{Soddu}},
  \bibinfo{journal}{Phys.Rev.} \textbf{\bibinfo{volume}{D72}},
  \bibinfo{pages}{095001} (\bibinfo{year}{2005}), \eprint{hep-ph/0509006}.

\bibitem[{\citenamefont{Aoki et~al.}(2000)}]{aoki}
\bibinfo{author}{\bibfnamefont{S.}~\bibnamefont{Aoki}} \bibnamefont{et~al.}
  (\bibinfo{collaboration}{JLQCD Collaboration}), \bibinfo{journal}{Phys.Rev.}
  \textbf{\bibinfo{volume}{D62}}, \bibinfo{pages}{014506}
  (\bibinfo{year}{2000}), \eprint{hep-lat/9911026}.

\bibitem[{\citenamefont{Giudice et~al.}(2011)\citenamefont{Giudice, Gripaios,
  and Sundrum}}]{Giudice:2011ak}
\bibinfo{author}{\bibfnamefont{G.~F.} \bibnamefont{Giudice}},
  \bibinfo{author}{\bibfnamefont{B.}~\bibnamefont{Gripaios}}, \bibnamefont{and}
  \bibinfo{author}{\bibfnamefont{R.}~\bibnamefont{Sundrum}},
  \bibinfo{journal}{JHEP} \textbf{\bibinfo{volume}{1108}}, \bibinfo{pages}{055}
  (\bibinfo{year}{2011}), \eprint{1105.3161}.

\bibitem[{\citenamefont{Adams}(2012)}]{Exotica}
\bibinfo{author}{\bibfnamefont{D.}~\bibnamefont{Adams}},
  \bibinfo{journal}{presented at 47th Rencontres de Moriond on Electroweak
  Interactions and Unified Theories, La Thuile, Italy}  (\bibinfo{year}{2012}).

\bibitem[{\citenamefont{Chatrchyan
  et~al.}(2011{\natexlab{b}})}]{Chatrchyan:2011ns}
\bibinfo{author}{\bibfnamefont{S.}~\bibnamefont{Chatrchyan}}
  \bibnamefont{et~al.} (\bibinfo{collaboration}{CMS Collaboration}),
  \bibinfo{journal}{Phys.Lett.} \textbf{\bibinfo{volume}{B704}},
  \bibinfo{pages}{123} (\bibinfo{year}{2011}{\natexlab{b}}),
  \eprint{1107.4771}.

\end{thebibliography}
\end{document}